\documentclass[final,authoryear,5p,times,twocolumn]{elsarticle}

\usepackage{amsmath}
\usepackage{amssymb}
\usepackage{lineno,hyperref}
\modulolinenumbers[5]

\newcommand*\patchAmsMathEnvironmentForLineno[1]{%
  \expandafter\let\csname old#1\expandafter\endcsname\csname #1\endcsname
  \expandafter\let\csname oldend#1\expandafter\endcsname\csname end#1\endcsname
  \renewenvironment{#1}%
     {\linenomath\csname old#1\endcsname}%
     {\csname oldend#1\endcsname\endlinenomath}}%
\newcommand*\patchBothAmsMathEnvironmentsForLineno[1]{%
  \patchAmsMathEnvironmentForLineno{#1}%
  \patchAmsMathEnvironmentForLineno{#1*}}%
\AtBeginDocument{%
\patchBothAmsMathEnvironmentsForLineno{equation}%
\patchBothAmsMathEnvironmentsForLineno{align}%
\patchBothAmsMathEnvironmentsForLineno{flalign}%
\patchBothAmsMathEnvironmentsForLineno{alignat}%
\patchBothAmsMathEnvironmentsForLineno{gather}%
\patchBothAmsMathEnvironmentsForLineno{multline}%
}

\usepackage{fixltx2e}

\journal{Journal of \LaTeX\ Templates}

\biboptions{authoryear}

\usepackage{color}

\begin{document}

\begin{frontmatter}

\title{V-shaped Cherenkov images of magnetically-separated gamma-rays}


\author[afil]{Julian Sitarek}
\ead{jsitarek@uni.lodz.pl}
\author[afil]{Dorota Sobczy\'nska}
\ead{dsobczynska@uni.lodz.pl}
\author[afil]{Katarzyna Adamczyk}
\ead{kadamczyk@uni.lodz.pl}
\author[afil2]{Micha\l\ Szanecki}
\ead{mitsza@camk.edu.pl}

\address[afil]{Department of Astrophysics, The University of \L\'od\'z, ul. Pomorska 149/153, 90-236 \L\'od\'z, Poland}
\address[afil2]{CAMK, ul. Bartycka 18, 00-716 Warsaw, Poland}

\begin{abstract}
Cherenkov Telescope Array (CTA) is an upcoming instrument that will start a new generation of atmospheric Cherenkov telescopes. 
CTA is expected not only to provide an unprecedented sensitivity in the tens of GeV to hundreds of TeV range, but also to considerably improve the systematic uncertainties of the measurements. 
We study the images registered by Cherenkov telescopes from low energy gamma rays with its first interaction in the upper parts of the atmosphere.
The images show a characteristic separation due to the deflection of the first $e^-e^+$ pair in the Geomagnetic Field. 
We evaluate the performance of the standard stereoscopic analysis for such events.
We derive also a novel method for energy estimation of V-shaped events based purely on geometrical properties of the image.
We investigate the potential of combining the classical energy estimation and the novel method for independent validation of the systematic shifts in the energy scale of Cherenkov telescopes and discuss the limitations of such analysis. 
\end{abstract}


\end{frontmatter}

\newcommand{\progname}[1]{{\fontfamily{pcr}\selectfont #1}}

\section{Introduction}

For the last 30 years the Cherenkov telescopes have been the main driver of studies of galactic and extragalactic TeV gamma-ray sources. 
The IACT (Imagining Air Cherenkov Telescopes) technique is based on the measurement of Cherenkov photons produced in the atmosphere by charged relativistic particles forming an Extensive Air Shower (EAS) initiated by the primary gamma-ray. 
The two-dimensional angular distribution of Cherenkov light appears on the telescope camera as the shower image. 
The parametrization \citep{hi85} of the image allows us to estimate the energy and direction of the primary particle, and to exclude a large fraction of the dominating hadronic background.
The rapid progress of both the hardware developments and new analysis methods resulted in great improvement in the sensitivity of Cherenkov telescopes (see e.g. Fig. 17 of \citealp{al16b}).
Currently, three major IACT instruments are in operation: H.E.S.S. \citep{aha06}, MAGIC \citep{al16a} and VERITAS \citep{weekes2002}. 
Already with the current generation of IACTs it is possible to obtain for bright flares precision measurements of nightly fluxes with statistical uncertainty of only a few per cent \citep[see e.g.][]{ah07,fu15,ah16}.  
Nevertheless, the Cherenkov telescopes are limited by relatively large systematic uncertainties, usually the dominating one is the uncertainty in the energy scale.
It is caused mainly by the uncertainty in the atmospheric production and absorption of the Cherenkov photons as well as lack of a ``test beam'' for IACT instruments. 
Despite various calibration methods (inter-telescope calibration: \citealp{ho03}, muon analysis: \citealp{iact_muons}, flux comparisons at different thresholds: \citealp{al16b}, LIDAR corrections: \citealp{fg15,de19}) it is still at the level of $\sim15\%$ (see e.g. \citealp{ah06, ab15, al16b}), much larger than of satellite experiments (c.f. 2\% energy scale systematic uncertainty estimated for \textit{Fermi}-LAT, \citealp{ack12b}). 
Such an uncertainty for a typical gamma-ray source can easily produce a $\sim30\%$ systematic error in the absolute normalization of the flux, with a more severe effect for steep spectrum sources and the energies close to the threshold. 

The upcoming Cherenkov Telescope Array (CTA) \citep{actis11,acha13} will start a next generation of large IACT arrays. 
CTA is designed to study gamma-ray sources from a few tens of~GeV to hundreds of~TeV with unprecedented sensitivity. 
To reach this goal it will be composed of sub-arrays of telescopes of different sizes: SST, MST and LST (small, medium and large sized telescope, respectively). 
To fully exploit the scientific potential of CTA, the increase in the statistical accuracy has to be matched also by efforts to lower systematic uncertainty of the measurements. 
In fact, various studies are being performed to lower the systematic uncertainty of the CTA: using atmospheric monitoring \citep{ga15}, muon calibration \citep{br15}, cosmic ray electron spectrum \citep{pa16}, inter-telescope calibration \citep{mi16}.

An interesting method to verify the energy calibration of ground instruments exploits the magnetic field of the Earth (i.e. Geomagnetic Field, GF) to measure the deflection of charged cosmic rays and the resulting shift of the shadow produced by the Moon or the Sun. 
Even while such a method proved successful for surface arrays of detectors like MILAGRO or HAWC \citep{chr11, ab13}, its application to IACTs is difficult due to strong background of moonlight. 
However, the Moon shadow measurements were performed with CLUE experiment, an array of Cherenkov telescopes sensitive only in UV range where the atmospheric absorption is high \citep{al98}.
Moreover, so far GF has been considered as a nuisance to the IACT technique, due to its effect on the images of the showers \citep[see e.g.][]{bo92, comm08, sz13}. 

In this work we investigate the feasibility of a novel method for verifying the energy calibration of CTA. 
The method is based on a search of characteristic images created by separation of the first $e^-e^+$ pair in the upper parts of the atmosphere by GF. 
In Section~\ref{sec:vevents} we perform basic analytical estimations to evaluate properties of such events. 
In Section~\ref{sec:sim} we describe the used simulation and analysis pipeline. 
Results of the simulations are presented in Section~\ref{sec:res}. 
The paper is concluded with discussion in Section~\ref{sec:con}.

\section{Magnetically separated events}\label{sec:vevents}
In order to evaluate what properties such events would have let us consider a $E_0\sim100$\,GeV gamma ray entering the atmosphere at a low zenith angle. 
On average after $\sim9/7$ radiation lengths, i.e. at the height of $H_0 \sim 22$\,km it will interact with atmospheric nuclei producing $e^-$ and $e^+$ with energies $E_-=f_{e^-}\,E_0$ and $E_+=(1-f_{e^-})\,E_0$ respectively. 
Initially $e^-$ and $e^+$ will be travelling nearly in the same direction as the primary gamma ray (the angle between the direction of $e^\pm$ and primary gamma rays will be of the order of arcsecond). 
However as they move through the atmosphere they will be more and more deflected by the GF, loosing at the same time energy by production of multiple Bremsstrahlung gamma rays. 
To estimate a characteristic deflection we can assume that the deflection occurs on a distance, $d_1$, corresponding to one radiation length, over which $e^+e^-$ will lose most of its energy. 
For the further calculations we will assume a simple exponential atmosphere with thickness $X = 1340 \cdot e^{-0.155 H/\mathrm{km}}/\cos\mathrm{ZD}\mathrm{\,[g\,cm^{-2}]}$\footnote{For close to zenith observations, this formula describes the La Palma atmosphere model used in standard CTA simulations with accuracy better than 5\% between height of 6 and 35\,km.}, where $\mathrm{ZD}$ is the zenith angle of the shower.
We obtain:
\begin{equation}
d_1=\frac{6.4\,\mathrm{[km]}}{\cos\mathrm{ZD}}\ln\left(1+\frac{X_0\cos\mathrm{ZD}}{1340 \mathrm{[g\,cm^{-2}]\exp(-0.155 H_1/\mathrm{km})}}\right),
\end{equation}
where $X_0=37\mathrm{[g\,cm^-2]}$ is the radiation length in air, and $H_1$ is the height of the first interaction. 
For $f_{e^-}\approx0.5$ we obtain the total deflection angle between $e^+$ and $e^-$ of:
\begin{equation}
\Psi_{+-} \approx 2d_1/R_L = 0.1^\circ (d_1/4\,\mathrm{km}) (E_0/100\,\mathrm{GeV})^{-1} (B_\perp/0.4\,\mathrm{G}),\label{eq:def}
\end{equation}
where $R_L$ is the Larmor radius of $e^+$/$e^-$, and $B_\perp$ is the component of GF perpendicular to the shower axis. 
Such a separation is within reach of the angular resolution of the CTA telescopes. 
In fact, the separation can be even more pronounced if the first interaction occurs higher in the atmosphere (resulting in longer $d_1$), and also taking into account that the primary $e^+$ and $e^-$ will loose energy during the deflection. 
Nevertheless, as most of the $e^+e^-$ energy is lost early, the Eq.~\ref{eq:def} is still a reasonable approximation of the effective deflection.  
The two subshowers started by $e^+$ and $e^-$ can be then in principle angularly separated. 
The angular separation of the two subshowers also shifts the positions of the Cherenkov photons on the ground, which can be roughly estimated as $\Psi_{+-}H_1/\cos\mathrm{ZD}\approx40$\,m for the above scaling values. 
Such a separation is a factor of a few below the radius of the shower lightpool ($\sim120$\,m). 
Hence we conclude that usually the same telescope would be able to see both subshowers, however propagating in a slightly different direction. 
Therefore such events observed under favourable conditions would produce characteristic V-shaped images. 
The difference of such V-shaped images to the magnetically broadened events studied e.g. by \citealp{bo92, comm08,sz13} should be stressed. 
These authors focus on the global effect of broadening of the image, mostly caused by the deflection of copious, lower energy $e^+e^-$ pairs produced in far generations in GF over shorter distances. 
Due to large number of individual $e^+e^-$ pairs with different energies and different distances from the shower axis, such broadening would not cause a clear separation of the image in two components, which naturally occurs in the scenario presented here. 
Note however, that each of the two sub-images in V-shaped events would be broadened by this effect, which is automatically taken into account in our simulations. 

\begin{figure*}[t]
\centering
\includegraphics[width=0.6\textwidth] {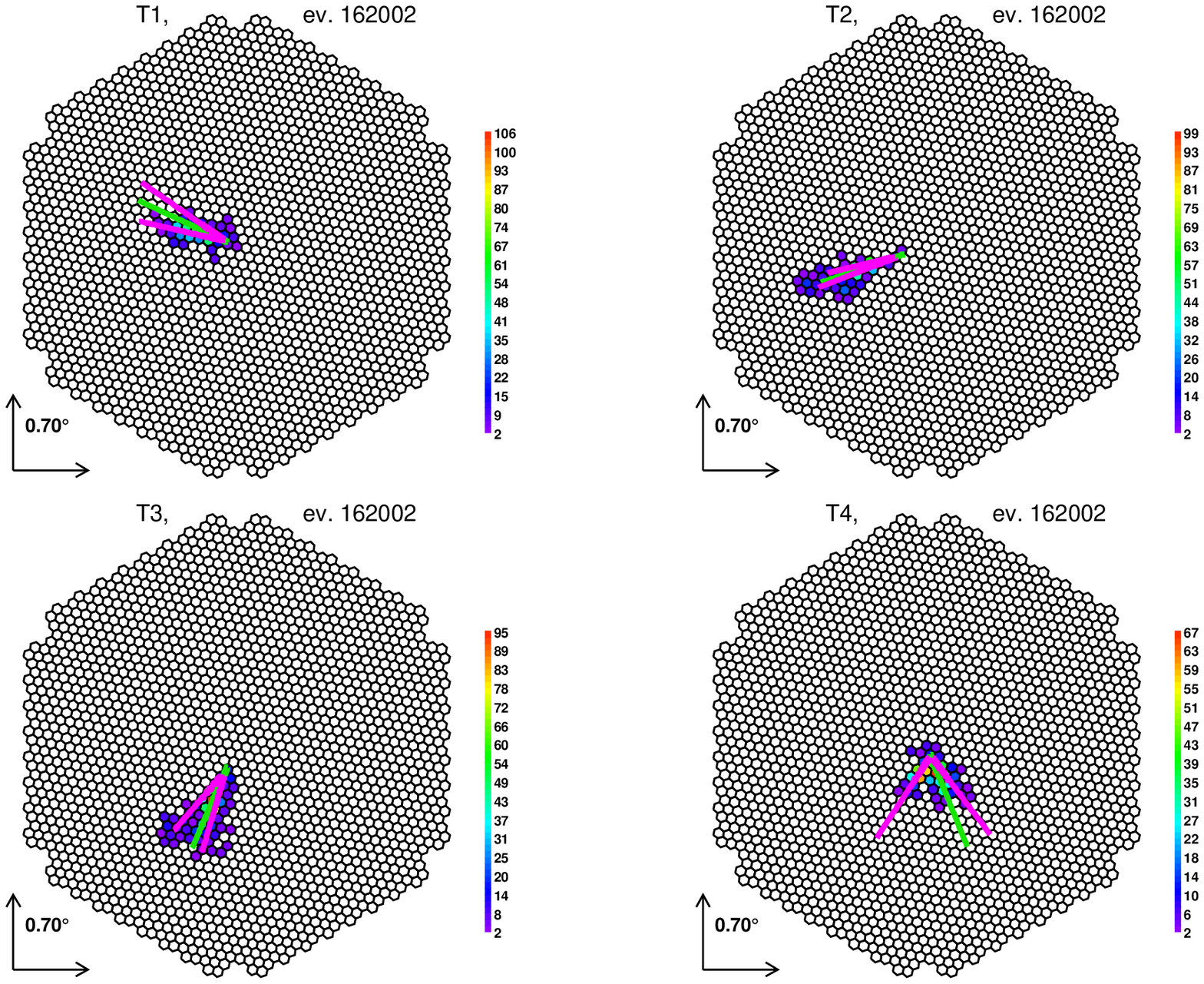}
\includegraphics[width=0.49\textwidth]{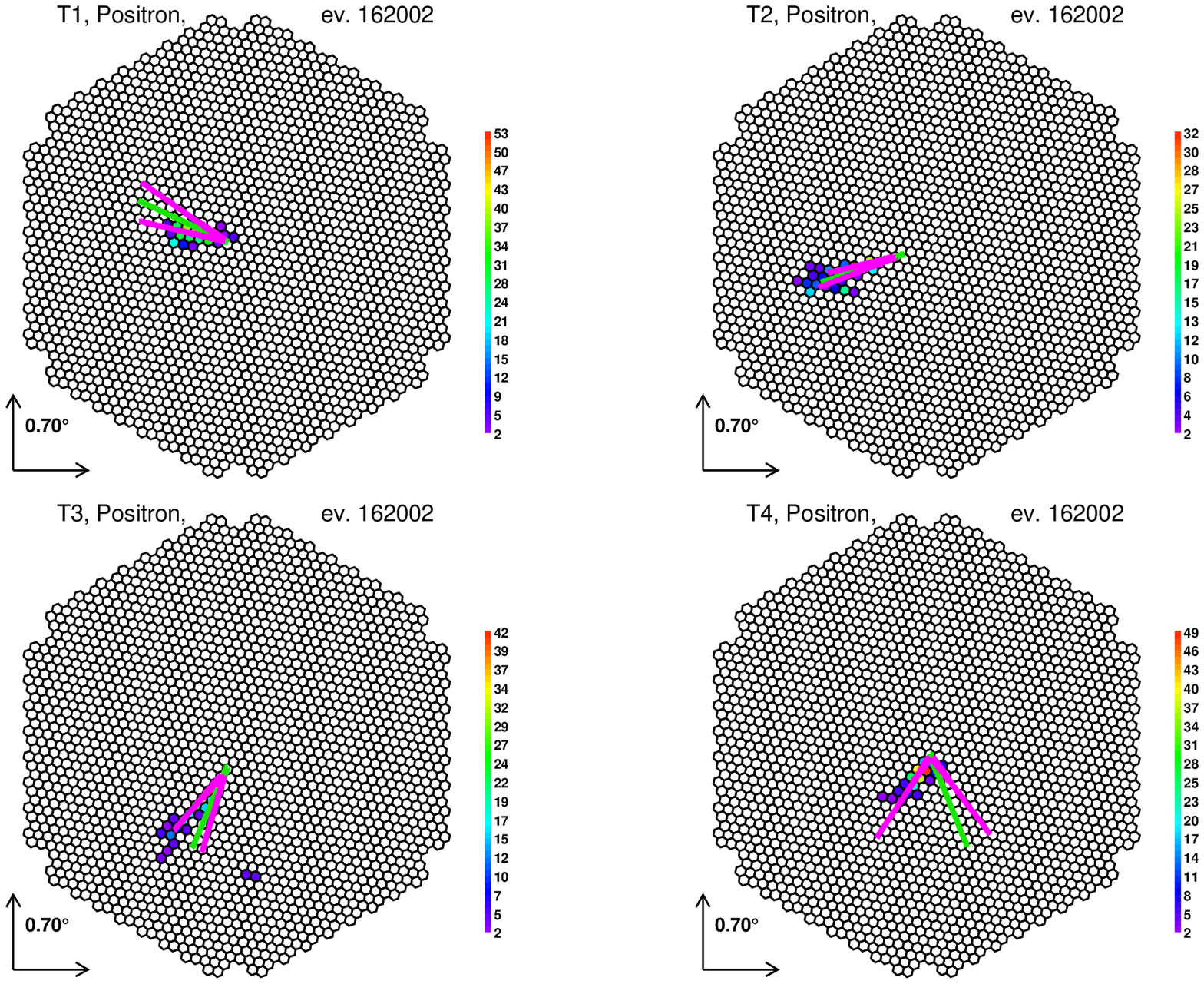}
\includegraphics[width=0.49\textwidth]{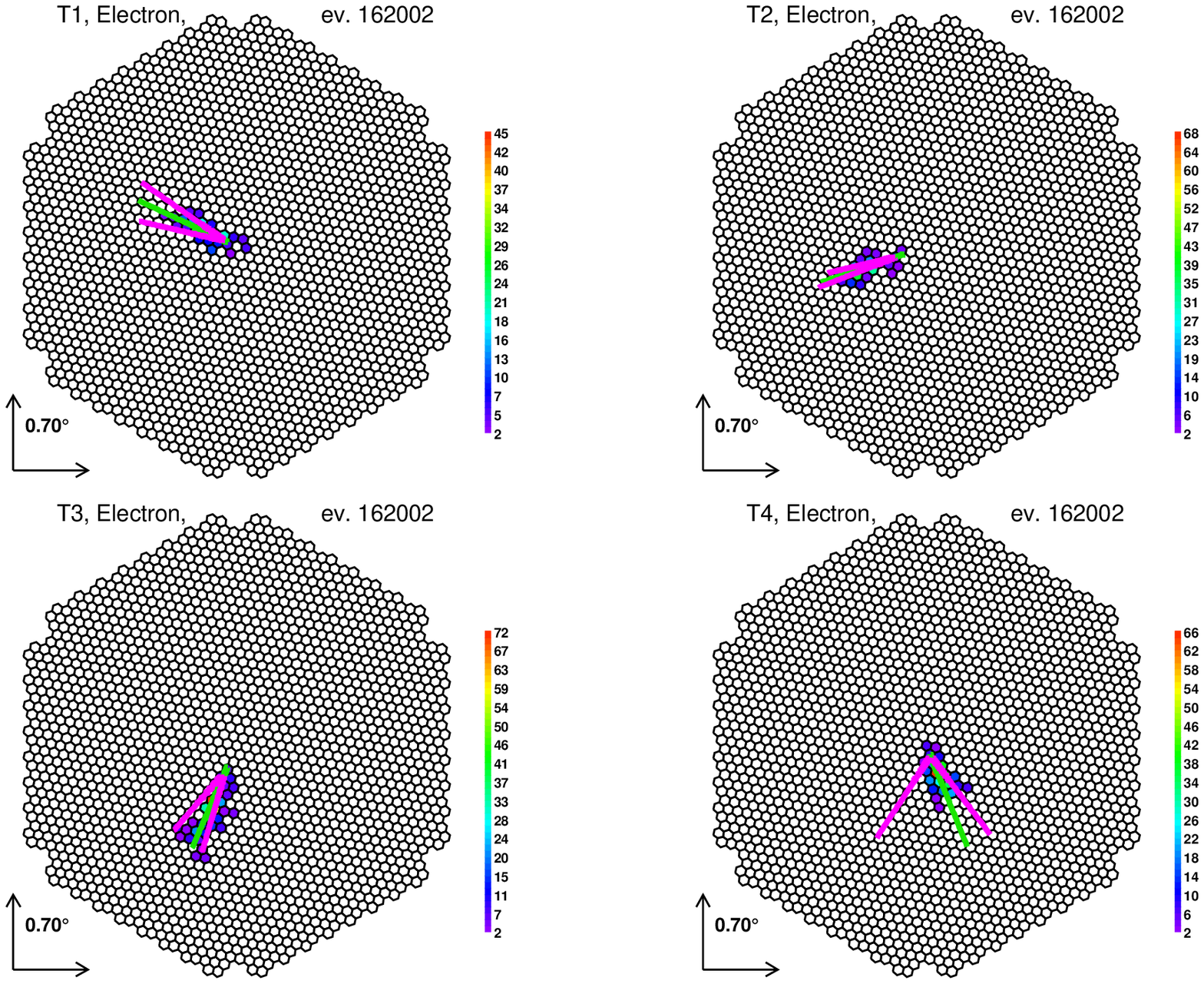}
\caption{Example of a V-shaped event as seen by 4 LST telescopes (different panels: T1, T2, T3, T4). 
The gamma ray with energy 186\,GeV had its first interaction at the height of 29.9\,km. 
The top 4 panels show the whole event, while the bottom panels show an image produced by subcascade initiated by $e^+$ (left) or $e^-$ (right) produced in the first interaction.
The color scale represents the number of phe reconstructed in each pixel. 
The lines show the predicted by the toy model (see Section~\ref{toy:model}) direction of images: the direction of gamma-ray shower if no separation would occur (green) and 
the direction of $e^+$ and $e^-$ subcascades (magenta). 
}\label{fig:image}
\end{figure*}

\begin{figure}[t]
\includegraphics[width=0.49\textwidth]{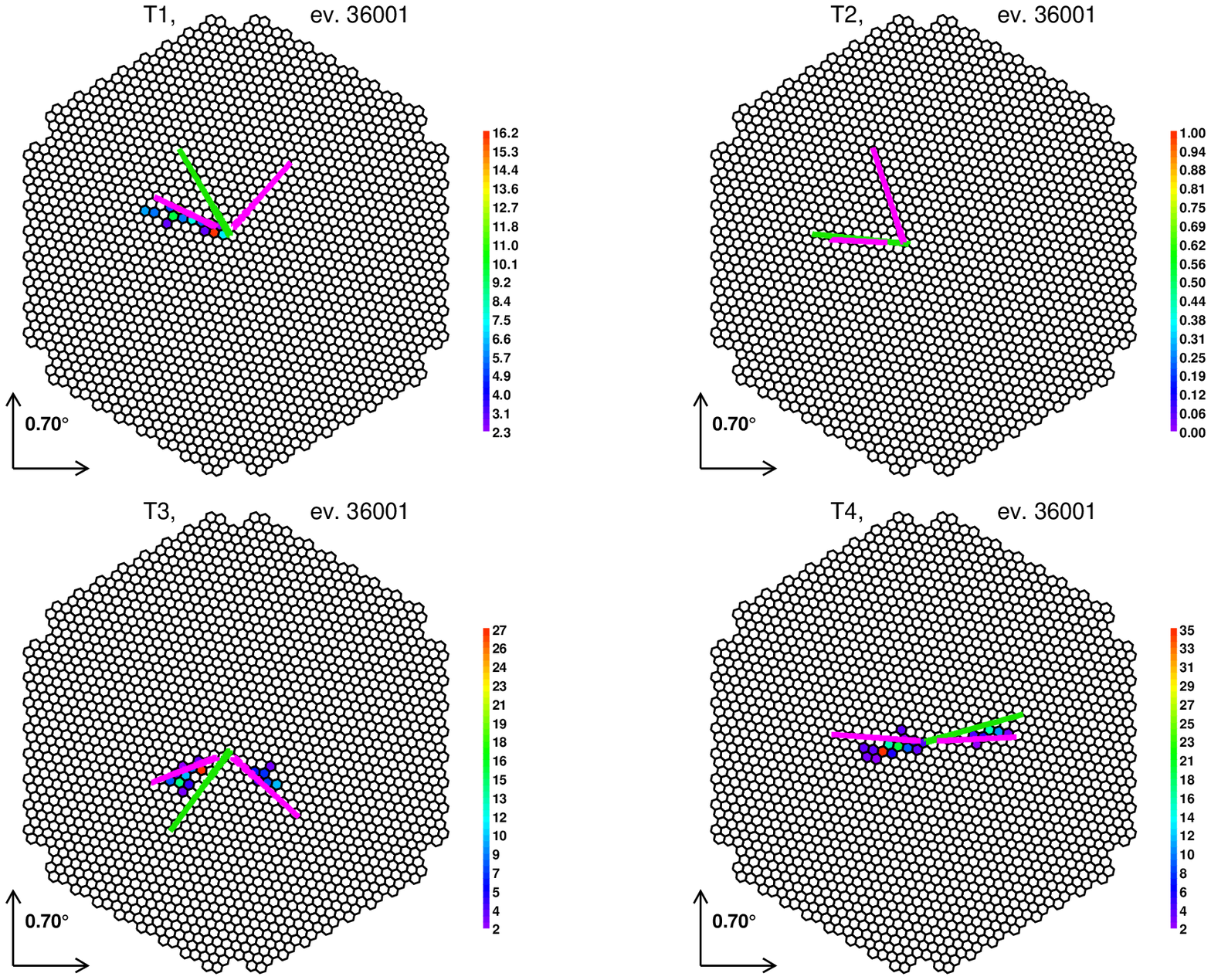}
\caption{As in the top panels of Fig.~\ref{fig:image}, however for a 69\,GeV shower with the first interaction at the height of 41.7\,km.  
}\label{fig:image2}
\end{figure}

In Fig.~\ref{fig:image} we show an example of an event with $e^+$ and $e^-$ subshowers separated as seen by 4 LST telescopes. 
In each pixel we plot the number of photoelectrons (phe) reconstructed from the signal recorded in that pixel. 
Night sky background and electronic noise is simulated independently in $e^+$ and $e^-$ and total images. 
The light produced in $e^+$ and $e^-$ subcascade is clearly shifted in the direction perpendicular to the magnetic field vector (i.e. for the simulated geometry in the horizontal direction in the camera). 
The total image (i.e. both the $e^+$ and $e^-$ subcascades together) of the shower in T4 (and to a lesser extent also in T2 and T3) is V-shaped. 
An even clearer separation can be achieved for lower energy showers and for showers that have started earlier in the atmosphere (see Fig.~\ref{fig:image2}). 
Note however that showers starting early in the atmosphere will produce dimmer images that are more difficult to reconstruct. 

\section{Simulations and data analysis}\label{sec:sim}
We perform simulations of gamma rays using \progname{CORSIKA} version 7.5 \citep{he98}.
We have simulated an array of 4 LST telescopes located in the La Palma site.
We simulate gamma rays with zenith angle of $20^\circ$ and azimuth corresponding to shower coming from geomagnetic North. 
Therefore the angle between the direction of the shower and GF is $72^\circ$, nearly maximizing the GF effects. 
The response of the telescopes was simulated using \progname{sim\_telarray} \citep{be08} using settings of the so-called \emph{CTA Prod 3 MC} \citep{acha19}. 
Both programs were modified to track additional information of gamma-ray induced showers.
Namely the particle type (most commonly $e^+$ and $e^-$ for a primary gamma ray shower) of the particles produced in the first interaction is propagated to the further generations of the cascade starting from those particles, and to the Cherenkov photons produced by the latter.
Therefore, for each emitted Cherenkov photon we know if it originated in the subcascade caused by $e^+$ or $e^-$ from the first interaction. 
The generated MC simulations are described in more detail in \cite{si18b}. 
The extraction of signal amplitudes from simulated waveforms, image cleaning (i.e. keeping only pixels with significant signals above NSB fluctuations) and its parametrization, gamma/hadron separation and classical stereo reconstruction and energy estimation is done using \progname{MARS/Chimp} chain \citep{za13, al16b, si18a}. 
Since the deflection of the two subshowers by the GF is roughly inversely proportional to the energy of the primary gamma we simulate only the LST subarray, where the strongest effect is expected. 

\section{Results}\label{sec:res}
In this section we present results of two types of simulations.
In the first, dedicated simulations, we exploit the extra knowledge of separation of shower into $e^+$ and $e^-$ subshowers to study how the separation into two subshowers affects the image.
Next we apply the V-shaped events selection cuts to the second, standard set of simulations, in which the composition of the image from $e^+$ and $e^-$ subshowers is unknown. 

\subsection{Comparison of $e^+$ and $e^-$ images}\label{sec:epem}
\begin{figure*}[t]
\centering
\includegraphics[width=0.33\textwidth]{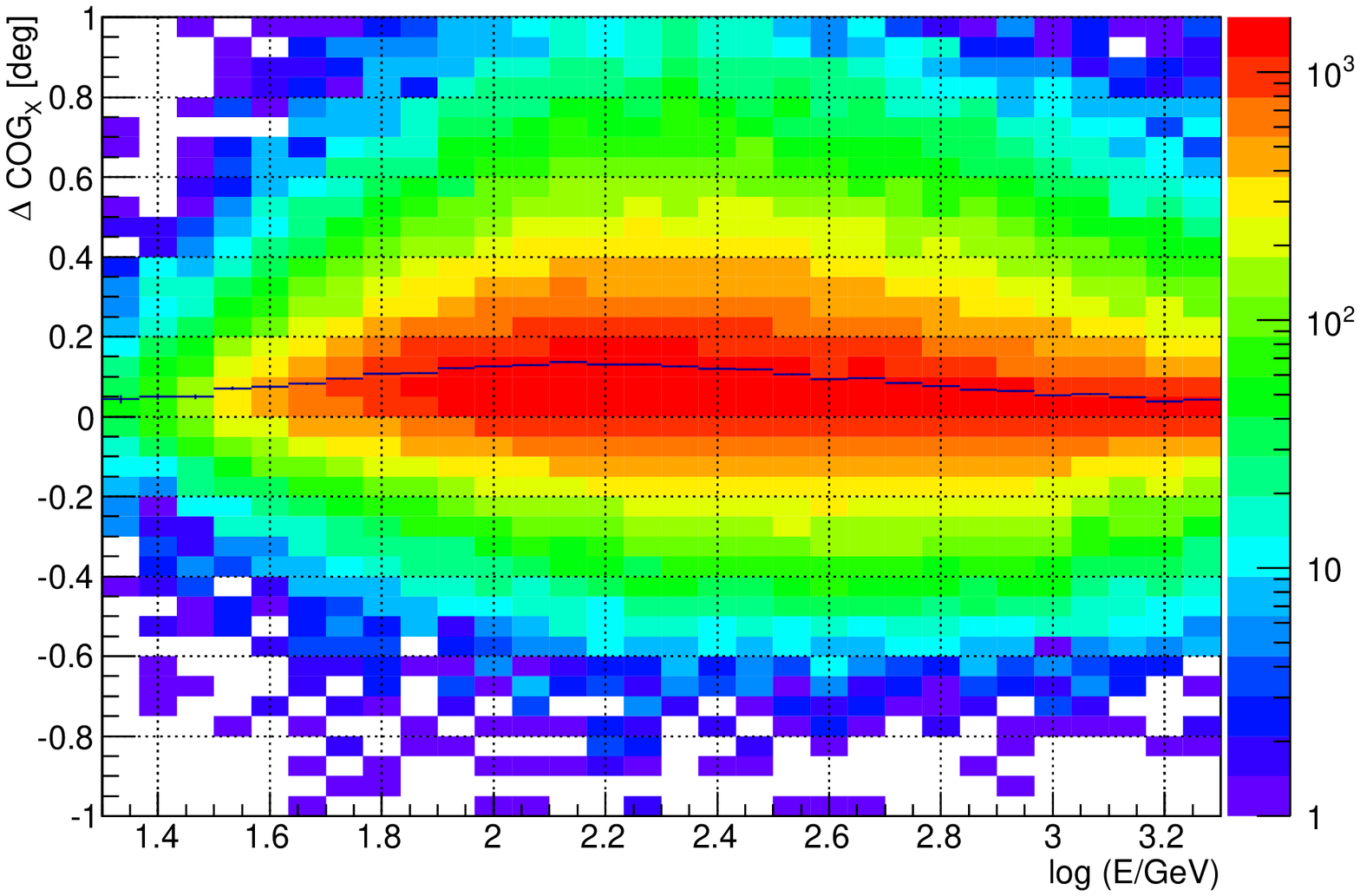}
\includegraphics[width=0.33\textwidth]{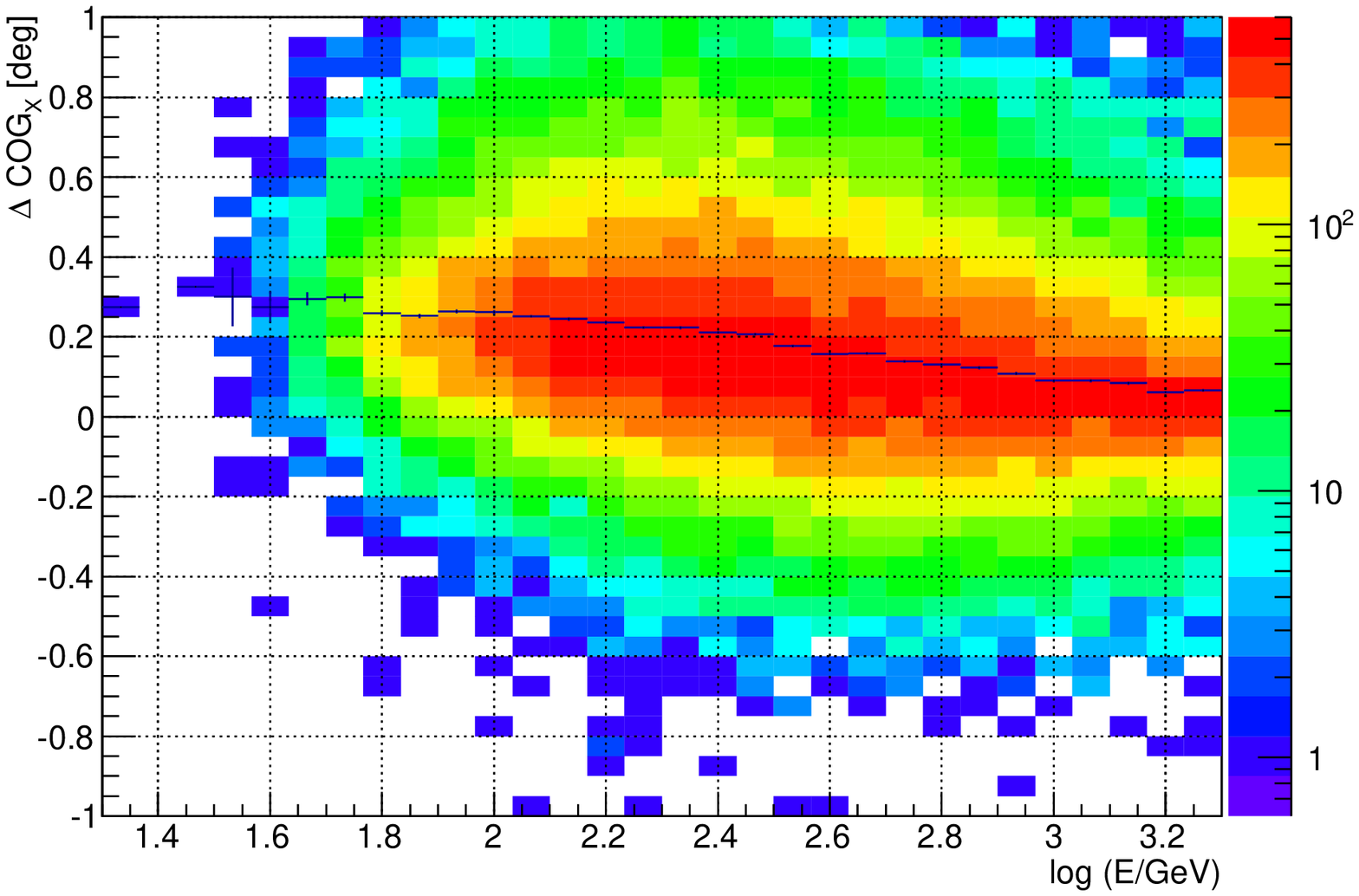}
\includegraphics[width=0.33\textwidth]{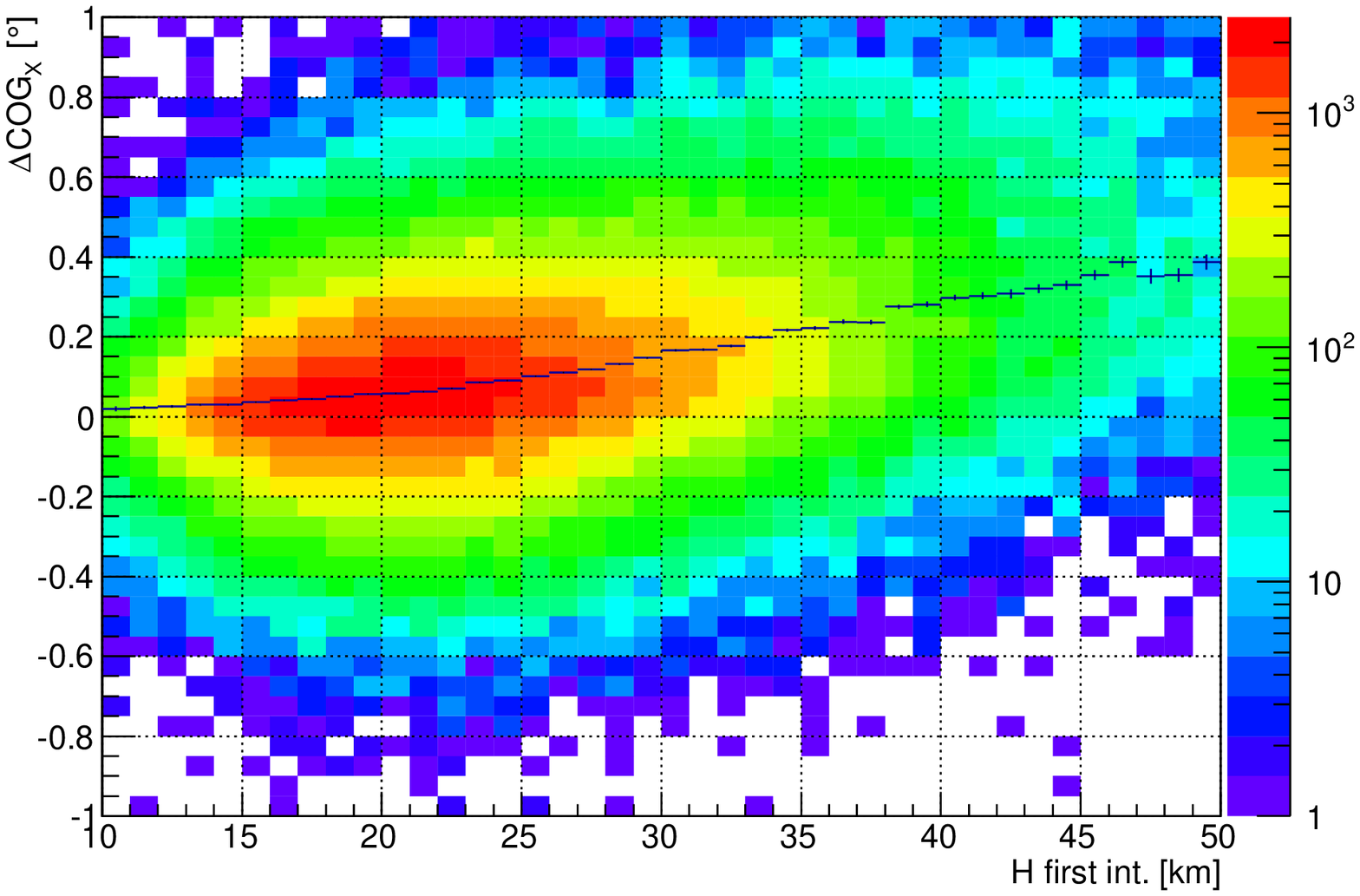}
\caption{Separation (measured perpendicularly to the projected direction of GF) of the images of the subshowers by the $e^+$ and $e^-$ produced in the first interaction of the primary gamma rays.
  The separation is plotted as a function of the energy of the gamma ray in the left panel.
  In the middle panel only events starting at the height above 25 km\, a.s.l. are shown.
  In the right panel the separation is plotted as a function of the height of the first interaction.
Black crosses show the profile of the histogram. 
Gamma rays coming from North are simulated with slope of $-2$. 
In all the panels only images in which the $e^+$ subcascade and $e^-$ subcascade produce a signal above 50\,phe each are plotted.
}\label{fig:dcog}
\end{figure*}

In order to investigate the separation effect of the $e^+$ and $e^-$ subcascades we analyze images composed by Cherenkov photons from only one of the two subcascades produced in the first interaction.
We clean separately images produced by $e^+$ and $e^-$ subcascades, calculate their Hillas parameters and exclude those in which at least one of the subcascades did not produce at least 50\,phe signal.
We define $\Delta\mathrm{COG_X}$ as the separation distance of the $e^+$ and $e^-$ subshower images, measured perpendicularly to the direction of the magnetic field projected onto the camera.
For the simulated here case of showers coming from geomagnetic North the positive value of $\Delta\mathrm{COG_X}$ means that the images are separated in the direction consistent with deflection in GF. 
In Fig.~\ref{fig:dcog} we investigate how $\Delta\mathrm{COG_X}$ depends on the energy of the primary gamma ray and its height of the first interaction. 
The image to image fluctuations are large, however a global shift of the distribution is clearly seen. 
At TeV energies the separation is small, $\sim0.05^\circ$. 
It increases going down in energy, to $\sim0.1^\circ$ at about 150~GeV, a consistent value with the result of the order of magnitude calculations of Eq.~\ref{eq:def}. 
Interestingly, at the lowest energies the average shift is getting smaller again (see the left panel of Fig.~\ref{fig:dcog}). 
This can be understood in terms of the dependence of the image separation on the height of the first interaction (the right panel of Fig.~\ref{fig:dcog}). 
Only those of the lower energy showers that are generated deeper in the atmosphere will allow enough light to be gathered by the telescopes for the triggering of the event and its consecutive reconstruction. 
Such events produced deeper in the atmosphere will be affected by a smaller deflection in the GF.
Limiting the calculations to only events starting high in the atmosphere ($>25$ km a.s.l., which at energy of $\sim100$\,GeV corresponds to roughly half of the events) shows a monotonic behaviour of the separation angle down to energies of ~30\,GeV (see the middle panel of Fig.~\ref{fig:dcog}).

\subsection{Selection of the v-images and rates}\label{sec:sel}
Since the images of low energy showers are composed of a small number of pixels, searching for two parts of an image by e.g. a fit with two two-dimensional Gaussian distribution, which has a large number of free parameters can lead to unstable, non-converging fits. 
Therefore, in order to preselect interesting images for more detailed analysis we apply, telescope-wise, a fast analytic ``double Hillas'' procedure. 
\begin{figure}[t]
\centering
\includegraphics[width=0.49\textwidth]{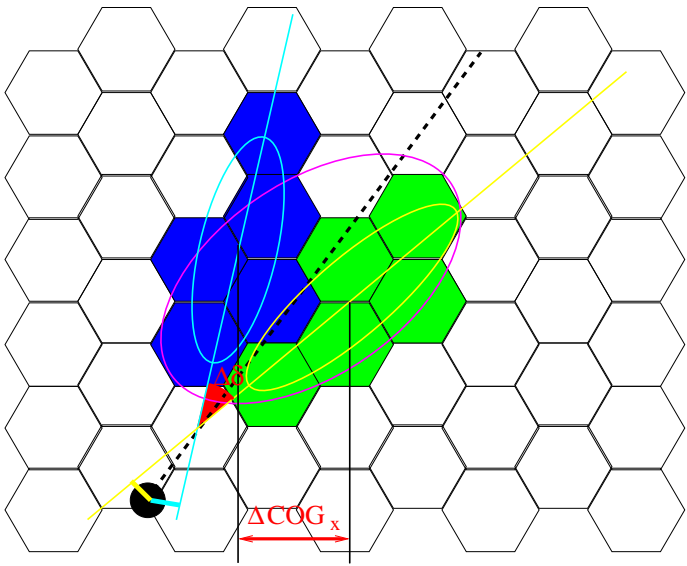}
\caption{
Sketch of double Hillas analysis. 
From pixels surviving the cleaning (filled hexagons) Hillas parameters are computed (magenta ellipse). 
Dashed line joins the source direction (black filled circle) and the COG of the whole image. 
The line divided the pixels into two groups (blue and green), from each Hillas ellipse is computed (cyan and yellow). 
The red angle $\Delta\delta$ is the angle between the main axes (thin blue and yellow lines) of the two ellipses on the camera.
Red double-headed arrow shows $\Delta\mathrm{COG_X}$, the distance of the COGs of the two ellipses measured perpendicularly to the direction of the GF (which is vertical in the assumed case). 
Short thick yellow and blue lines close to the circle show the \emph{Miss} parameter. 
}\label{fig:doublehillas}
\end{figure}
First we calculate the Hillas parameters of the full image. 
Next, the image is divided by a line crossing the source position on the camera and the COG of the image.
The pixels forming the image are divided into two groups depending on which side of the line their center is, and the Hillas parameters are calculated independently for each of the groups. 
If the image is a classical ellipse such a procedure will approximately cut it along its main axis resulting in two parts of the image separated by $\Delta\mathrm{COG_X}\lesssim 0.1^\circ$ (of the order of the Width parameter of the image).
The average angle between their main axes can be estimated as a ratio of $\Delta\mathrm{COG_X}$ to \emph{Dist} (the angular distance of the COG of the image from the direction of the primary gamma ray), i.e. $\Delta\delta\sim\arcsin(0.1^\circ/1^\circ)\approx 6^\circ$. 
On the other hand, in the case of images like T4 in Fig.~\ref{fig:image} or Fig.~\ref{fig:image2} double-Hillas analysis reveals the two components and results in a larger separation of COGs and main axes of the two parts of the image. 
To reject fake islands in the image due to night sky background remaining after the cleaning and also to limit the contribution of the hadronic background, we keep only the events in which both parts of the image separately are pointing towards the true source position. 
We achieve it by cutting in the so-called \emph{Miss} parameter, i.e. the distance between the main axis of the subshower image and the direction of the source.

\begin{figure}[t]
\centering
\includegraphics[width=0.49\textwidth]{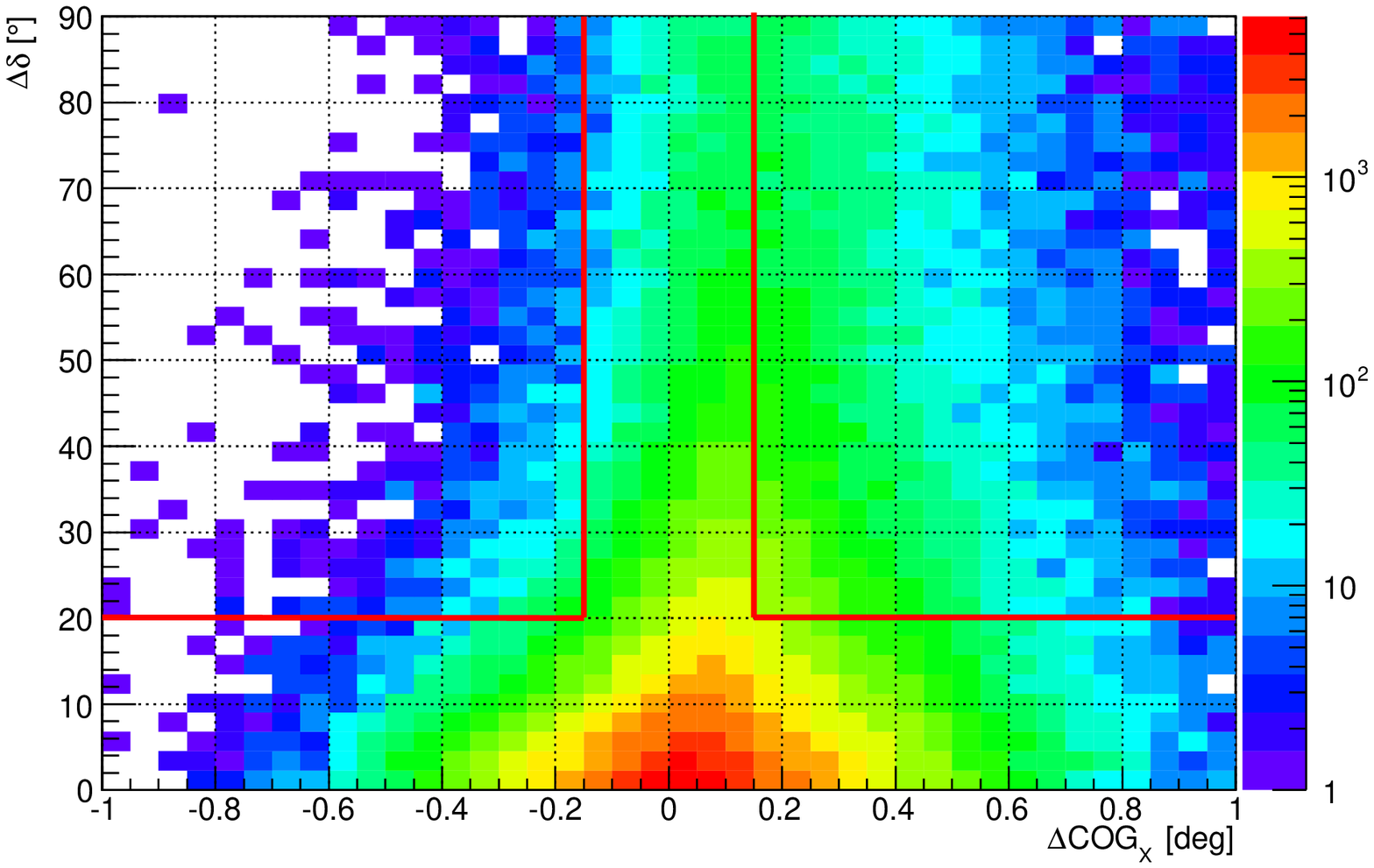}
\includegraphics[width=0.49\textwidth]{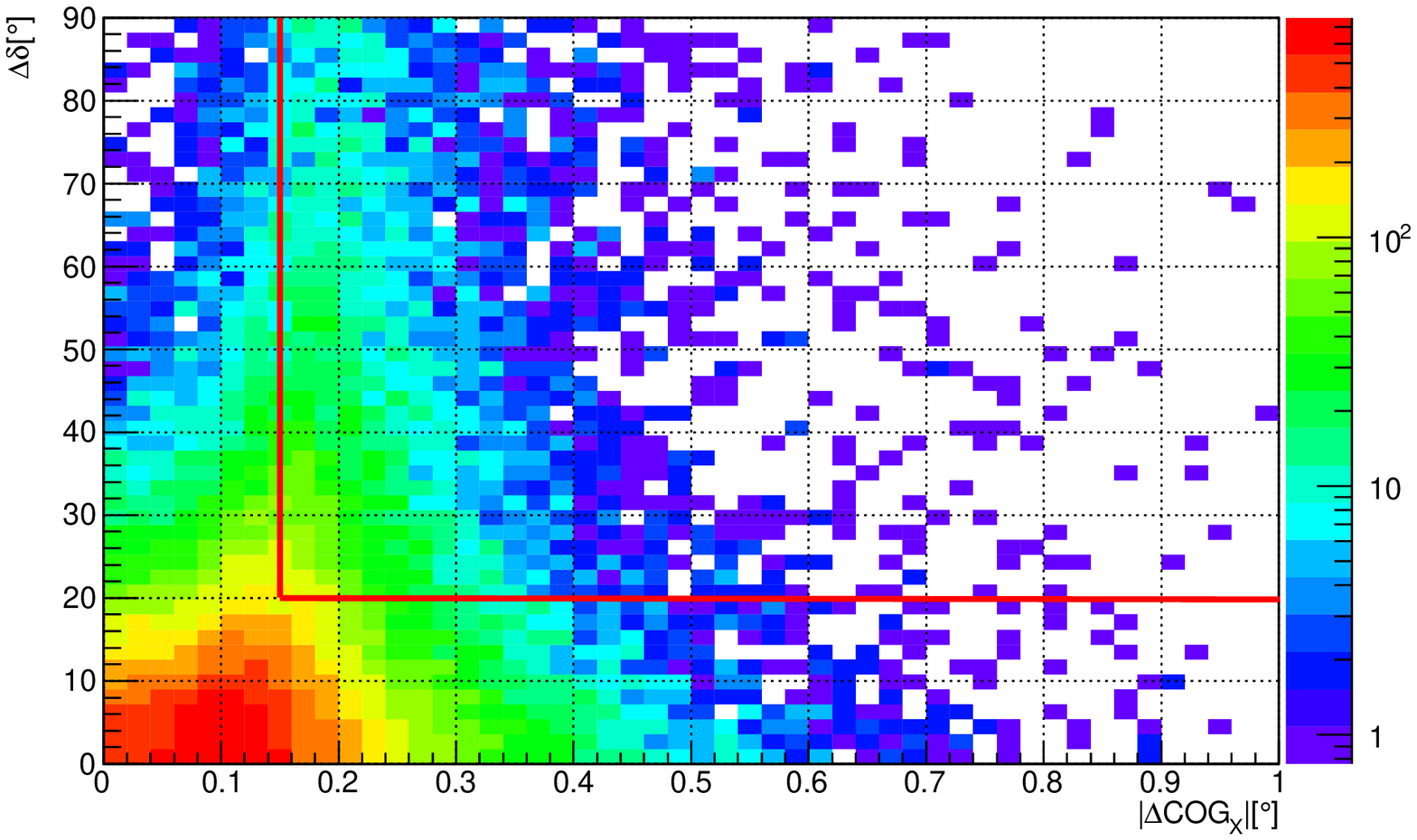}
\caption{
Top panel: angle between the main axis of the $e^-$ and $e^+$ subcascade images measured on the camera as a function of the distance between the COG of those images. 
Only images in which the $e^+$ subcascade and $e^-$ subcascade produce a signal above 50\,phe each, and both point in the direction of the source ($Miss_{Img1}<0.15^\circ$, $Miss_{Img2}<0.15^\circ$) are plotted.
Thick red lines show the values of the cuts $|\Delta\mathrm{COG_X}|>0.15^\circ$, $\Delta\delta>20^\circ$. 
Gamma rays coming from North are simulated with slope of $-2$. 
Bottom panel: same as above, however for the two parts of image from double-Hillas analysis (see the text for details). 
}\label{fig:cogvsdelta}
\end{figure}

In order to check the viability of the procedure described above  and to tune the values of the cuts in $\Delta\mathrm{COG_X}$ and $\Delta\delta$ in Fig.~\ref{fig:cogvsdelta} we compare the 
$\Delta\delta$ vs $\Delta\mathrm{COG_X}$ plot obtained from special MCs where $e^+$ and $e^-$ subcascades were processed separately (top panel) with the one obtained from regular MCs in which double-Hillas analysis was used. 
Note that in the latter case the subimages from $e^+$ and $e^-$ subcascades cannot be distinguished, i.e. only absolute value of $\Delta\mathrm{COG_X}$ can be derived. 
As shown already in Fig.~\ref{fig:dcog}, due to the deflection in GF, $\Delta\mathrm{COG_X}$ is shifted towards positive values. 
The most clear V-shaped events are located in the top right part of the top panel of Fig.~\ref{fig:cogvsdelta} (i.e. large $\Delta\delta$ and large positive $\Delta\mathrm{COG_X}$). 
The events with large $\Delta\delta$ but large negative $\Delta\mathrm{COG_X}$ are events in which large fluctuations of the shower mimick the separation due to GF deflection.
As the number of events with $\Delta\delta>20^\circ$ and $\Delta\mathrm{COG_X}<-0.15^\circ$ is much smaller than the number of events for which $\Delta\delta>20^\circ$ and $\Delta\mathrm{COG_X}>0.15^\circ$, we conclude that the background of such mimicking events will be small. 
The bottom panel of Fig.~\ref{fig:cogvsdelta} obtained with double-Hillas analysis of full images show a similar shape as the folded along the $\Delta\mathrm{COG_X}=0$ axis upper panel. 
As expected the classical elliptical events are located at small values of $\Delta\mathrm{COG_X}$ and $\Delta\delta$, while the more interesting, V-shaped events will produce larger values of $\Delta\mathrm{COG_X}$ and $\Delta\delta$. 
We define image as V-shaped if it fulfills the following conditions: $Miss_{Img1}<0.15^\circ$, $Miss_{Img2}<0.15^\circ$, $\Delta\mathrm{COG_X}>0.15^\circ$ and $\Delta\delta>20^\circ$.
$Miss_{Img1}$ and $Miss_{Img2}$ are the $Miss$ parameters calculated independently for the two separated by the double-Hillas analysis parts of the image. 

\begin{figure}[t]
\includegraphics[width=0.49\textwidth]{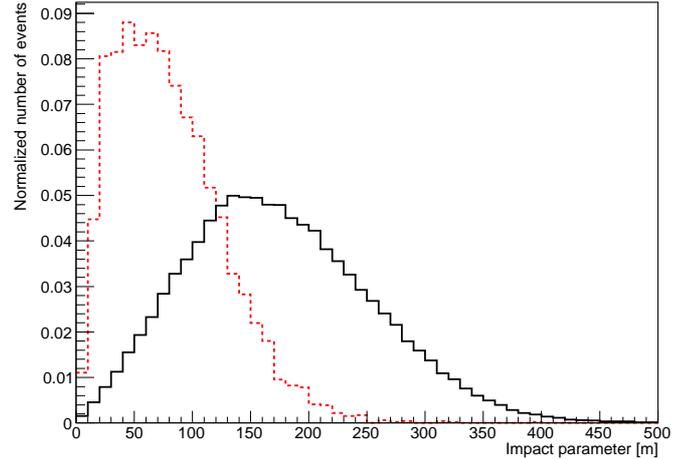}
\caption{
  Normalized distribution of the true impact parameter for images that fulfill the V-shaped criterion (red dotted) and for remaining images (black). 
  Only events with estimated energy between 30 and 300\,GeV are used. }\label{fig:impact_dist}
\end{figure}

In Fig.~\ref{fig:impact_dist} we compare the distribution of the true impact parameter (calculated with respect to each telescope) for V-shaped and not-V-shaped images. 
For V-shaped images the distribution is shifted to lower values and peaks around 60\,m.
It is caused by the geometrical effect that the same separation angle $\Psi_{+-}$ between the $e^+$ and $e^-$ subshowers for different impact parameters will result in different $\Delta\delta$ affecting also $\Delta\mathrm{COG_X}$ angle.
For example if the impact parameter is zero, the two subshowers will propagate in opposite direction on the camera ($\Delta\delta=180^\circ$), maximizing also $\Delta\mathrm{COG_X}$. 

\begin{figure}[t]
\includegraphics[width=0.49\textwidth]{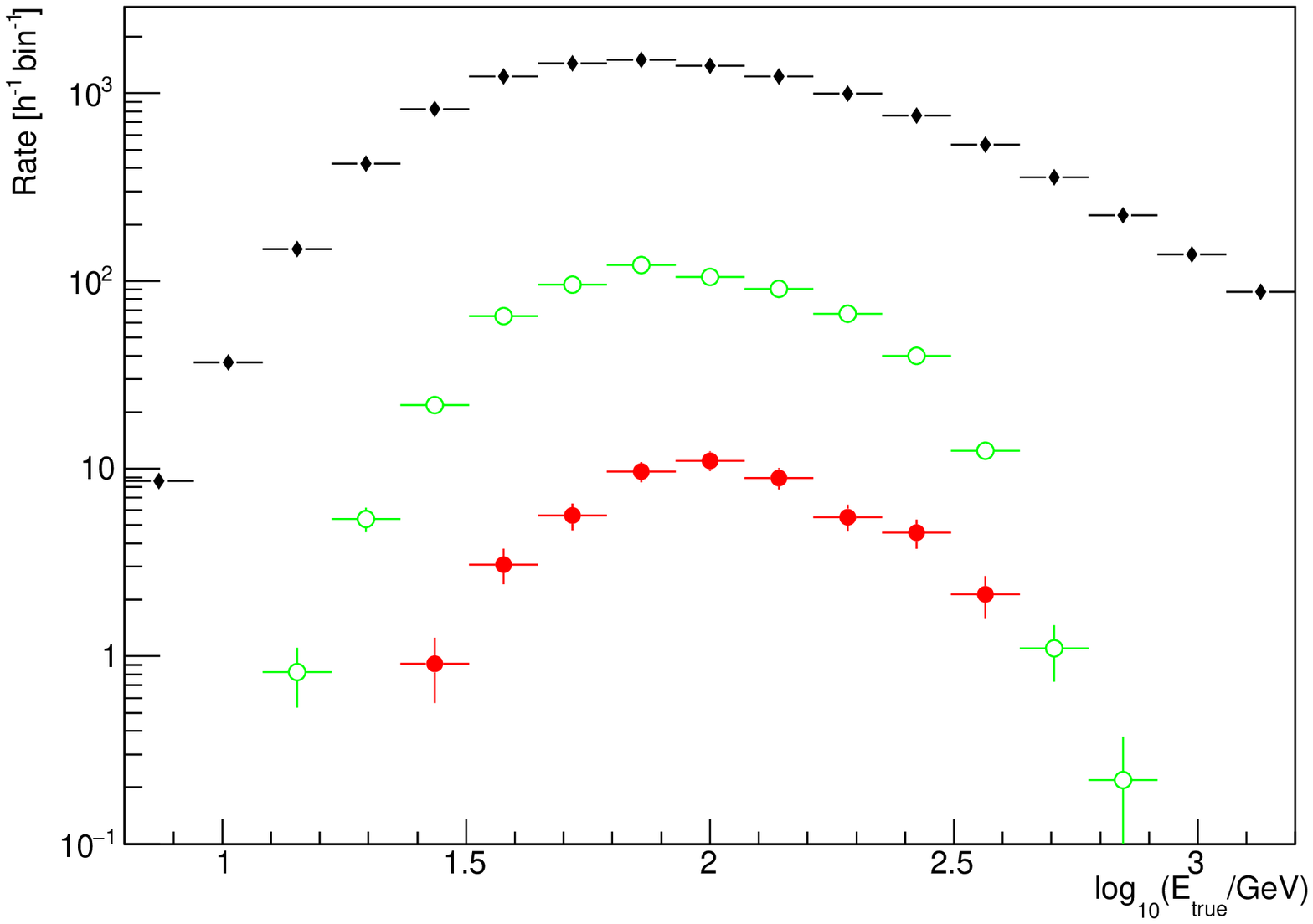}
\caption{
  Rate of events surviving V-shape selection cuts in at least one (green open circles) or at least two (red filled circles) images from a source with a Crab-like spectrum.
  Only events with estimated energy between 30 and 300\,GeV are used. 
  For comparison, the total rate of reconstructed gamma rays is shown as black diamonds.}\label{fig:ratev}
\end{figure}

In Fig.~\ref{fig:ratev} we show the rates of such preselected events in which at least one/two images are classified as V-shaped.
The rates are normalized to a source with a Crab-like spectrum (including its curvature following \citealp{al15}).
The lowest energies produce too dim images for both components to be clearly visible.
On the contrary, the highest energies produce large images in which the $e^+$ and $e^-$ subcascades are not-separatable.
Therefore we limit the analysis to events with estimated energy between 30 and 300\,GeV. 
At 100\,GeV about 1/10 of the events show at least one image with separation of the two components in the double Hillas analysis larger then the applied cuts.
At the same energy about 1/100 of the events show a clear indication of being V-shaped in at least two individual telescope images.
For observations of a source with a Crab like spectrum the total rate of the V-shaped events surviving the selection criteria in at least two images is 50 per hour.
For the further analysis we consider events with V-shape images present in at least two LST telescopes.

\subsection{Toy model}\label{toy:model}
According to Eq.~\ref{eq:def} the deflection of the two images depends on the energy of the primary gamma ray.
The classical energy estimation is based mainly on the amount of Cherenkov light for the reconstructed stereoscopic parameters (the impact parameter and the height of the shower maximum). 
Hence, comparing the classical energy estimation of those events with the developed here independent geometrical energy estimation has a potential for validation of the total light throughput of the telescopes and this in turn validates the energy calibration. 
Unfortunately $\Psi_{+-}$ depends also on the height of the first interaction and on the separation of the primary energy into $e^+e^-$. 
For individual events those are poorly known, however if a large sample of V-shaped events is gathered, this problem can be dealt with on a statistical basis. 

In order to apply such a calibration procedure to Cherenkov telescopes, precise enough knowledge of the GF is needed. 
In fact, the standard simulation program of EAS, \progname{CORSIKA}, has a somewhat simplified treatment of the GF, neglecting the height dependence. 
Nevertheless, according to the WMM2015 model\footnote{\url{https://www.ngdc.noaa.gov/geomag/WMM/}} the change of the GF components in our range of interest, i.e. between 10 and 40 km a.s.l. is only about 1.5\% for the La Palma site.

In order to test the potential of V-shaped events to calibrate/validate the energy scale of LST we developed a simple geometrical model of such events. 
The model assumes a knowledge on the direction of the shower $\vec{v}_0$ (azimuth and zenith angles: $\phi, \theta$), and local GF, $\vec{B}$, and location of all the telescopes (including relative height differences). 
In addition to this, it has 5 free parameters: the core position on the ground $(x_0, y_0)$, the height of the first interaction $H_0$, energy of the gamma ray $E_0$, and the fraction $f_{e^{-}}$ of the energy taken by the first generation $e^-$. 
All of the parameters (except of $f_{e^{-}}$) can be left free to vary in the fitting procedure or it can be estimated from other methods. 
In particular the core position can be estimated using stereoscopic reconstruction of the shower \citep{ho99}. 
The height of the first interaction, even while poorly constrained, can be estimated with the closest pixel method \citep{si18b} or with a model analysis \citep{nr09,par14}. 

Using $\vec{v}_0$, $(x_0, y_0)$ and $H_0$ we calculate first the point in space, $P_0$, where the first interaction happens. 
Starting from $P_0$ the first generation $e^+$ and $e^-$ pair diverges. 
We assume that each of the leptons traverses one radiation length $d_1$, which geometrical length is determined from $H_0$ and $\theta$ using a simple exponential profile of the atmosphere. 
The deflection angle of $e^+$ and $e^-$ is computed as $\Psi_+=d_1/R_{L,+}$ and $\Psi_-=d_1/R_{L,-}$ respectively, where $R_{L,+}$ and $R_{L,-}$ are their Larmor radii computed using energies $(1-f_{e^{-}})E_0$ and $f_{e^{-}}E_0$ and perpendicular part of $\vec{B}$.
We calculate the directions of the deflection of the $e^+$ and $e^-$ subshowers as the unit vectors
$\vec{n}_\pm = \pm\frac{\vec{v}_0\times\vec{B}}{|\vec{v}_0\times\vec{B}|}$.
The direction of the $e^\pm$ subshowers is then computed as
$\vec{s}_\pm = \frac{\vec{v}_0 + \Psi_\pm \vec{n}_\pm}{|\vec{v}_0 + \Psi_\pm \vec{n}_\pm|}$. 
The starting points of the $e^\pm$ subshowers are then assumed as
$P_\pm = P_0 + 0.5(\vec{s}_\pm+ \vec{v}_0) d_1$. 
Each subshower propagating from point $P_\pm$ along $\vec{s}_\pm$ direction will then produce a line in the coordinates system on the Cherenkov telescope camera. 
Therefore, for a given set of starting parameters we obtain 2 lines per each telescope. 

It should be noted, that the toy model does not take into account several effects. 
In particular the stochastic behaviour of Bremsstrahlung, energy losses of the first $e^\pm$ pair affecting the deflection, lateral distribution of the electrons in the subshowers and selection bias of the events triggered by the telescopes can modify the resulting image.

\subsection {Standard stereoscopic reconstruction for V-shaped events}
\begin{figure}[t]
\includegraphics[width=0.49\textwidth]{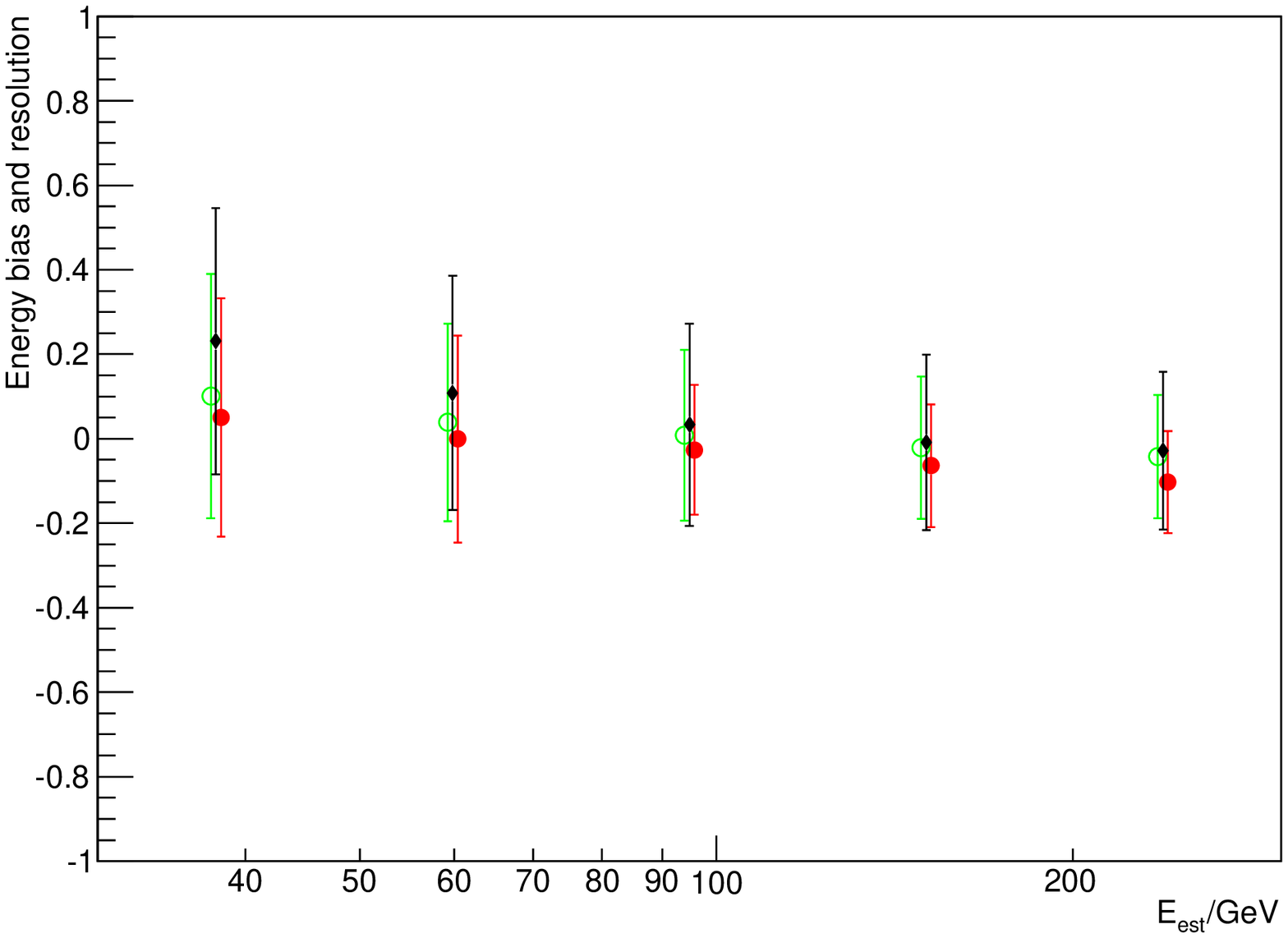}
\includegraphics[width=0.49\textwidth]{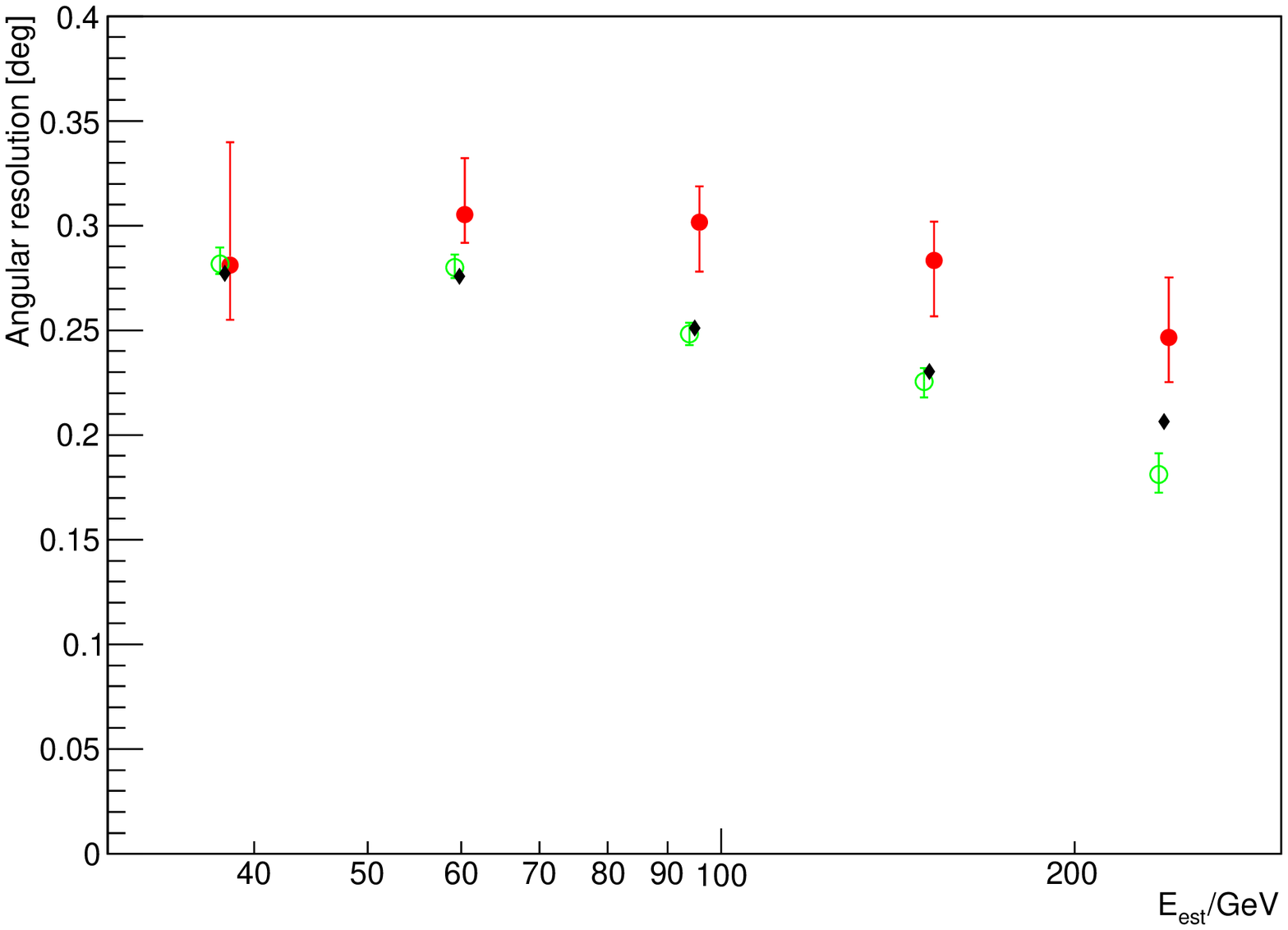}
\caption{Top panel: energy bias (point position) and resolution (represented as error bars) as a function of the estimated energy of the shower.
  Only events reconstructed within $0.1^\circ$ from the nominal source position are taken into account.
  Bottom panel: 68\% PSF containment radius angular resolution (and its uncertainty depicted as error bars) as a function of the estimated energy of the shower.
  Different colours show events of different types: with one V-shaped image (green open circles ), with at least two V-shaped imaged (red filled circles), no V-shaped images (black diamonds).
  All events that survive image reconstruction and energy estimation 
  with estimated energy 30--300 GeV are used.
  No cut in gamma/hadron separation parameter is applied. 
  For visibility the different curves are moved by $\pm$1\% in X direction.
  }
\label{fig:en_ang_res}
\end{figure}

In order to evaluate how the standard stereo reconstruction deals with the V-shaped events we compare the basic performance parameters for the events with (a) two or more, (b) one and (c) no V-shaped images.
Note that the simulations were performed for showers coming from North, 
where the effect of GF worsens the performance of the Cherenkov telescopes (see e.g. \citealp{ha17}). 
In Fig.~\ref{fig:en_ang_res} we show the comparison of angular and energy resolution for different classes of events.
The energy resolution we calculate as a second central moment of the distribution of $(E_{est}-E_{true})/E_{true}$, i.e. including non-Gaussian tails.
It should be noted that for those comparisons we do not apply any additional quality cuts (e.g. in number or telescopes used in the reconstruction), nor cuts in the  gamma/hadron separation parameter. %
Such cuts would improve the absolute values of the performance parameters.
For the calculation of the energy resolution we apply however a cut in the reconstructed position of the source. 
The performance parameters for events with some of the images V-shaped are comparable to events with classical elliptical images in all the telescopes.
This can be understood taking into account that the most clear V-shaped images occur if the impact parameter of the shower to the telescope is small (compare also Fig.~\ref{fig:impact_dist}) and both subcascades develop on opposite sides of the telescope.
In such a case the event is normally relatively well contained within the array of 4 LST telescopes and there are enough non-V images for efficient reconstruction.
Moreover, V-shaped images will contribute with lower weight in the energy reconstruction due to their unusual values of Hillas parameters (c.f. the \emph{Chimp} energy estimation procedure explained in \citealp{si18a}). 

\begin{figure}[t]
\includegraphics[width=0.49\textwidth]{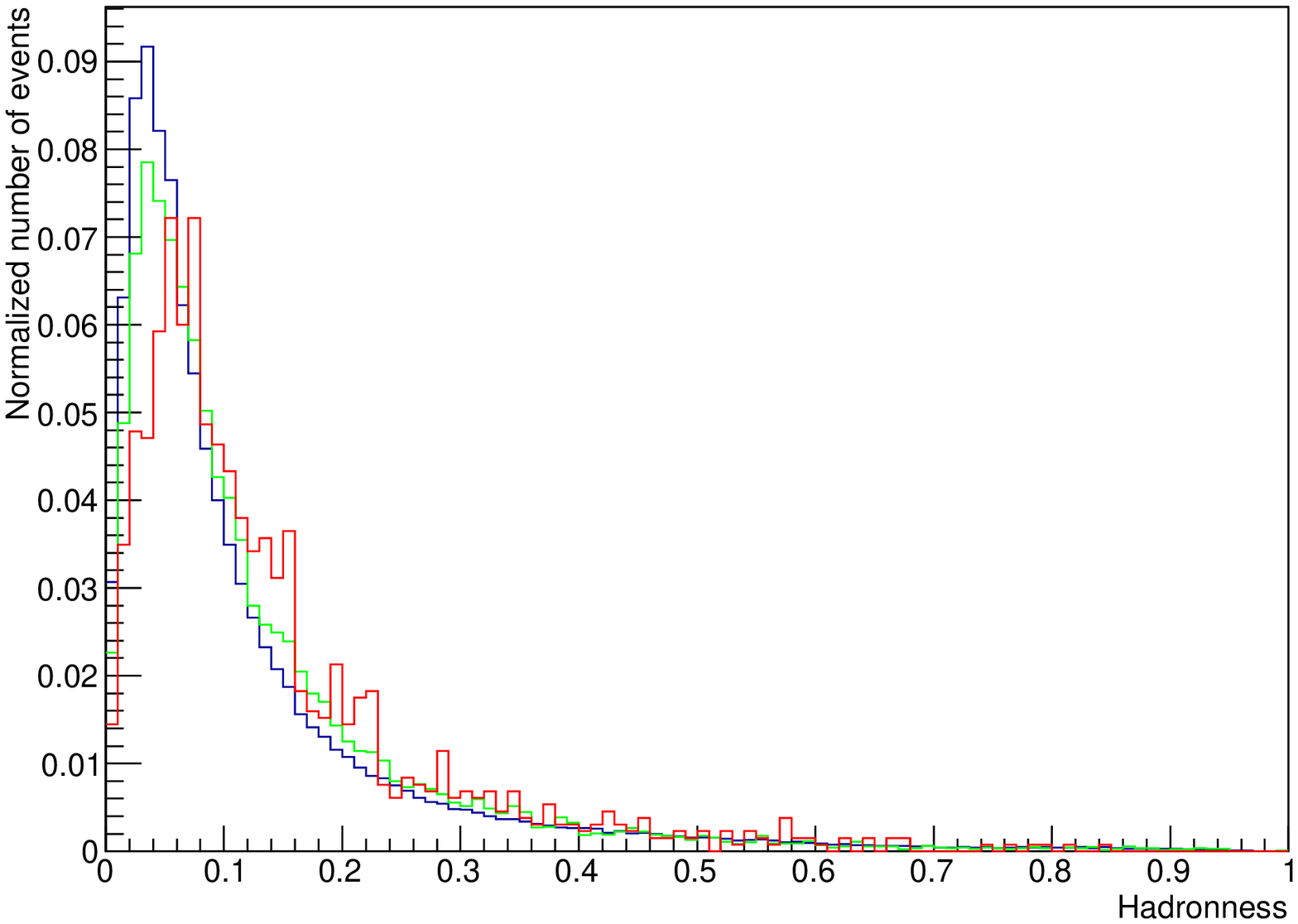}
\caption{Distribution of \emph{Hadronness} parameter for events with at least two V-shaped images (red), one V-shaped image (green), and no V-shaped events (dark blue). 
Only events with estimated energy between 30 and 300\,GeV are used.}
\label{fig:hadr}
\end{figure}

In Fig.~\ref{fig:hadr} we show the distribution of the gamma/hadron separation parameter \emph{Hadronness} for different classes of events.
V-shaped events when parameterized as Hillas ellipses have larger \emph{Width}, and in case of $\Delta\delta>90^\circ$ also larger \emph{Length} making them more hadron-like.
Therefore the distributions for events with V-shaped images are shifted towards larger \emph{Hadronness} values. 
Nevertheless the effect is small even for the case of two V-shaped images in one event.
The value of \emph{Hadronness} cuts that keeps 90\%, 80\% and 60\% of the events with no V-shaped images keeps also respectively 87\%, 74\% and 46\% of events with at least two V-shaped images. 
The fact, that both the angular resolution and the hadronness distribution are not much affected by occurance of the V-shaped images, allows one to use the standard analysis methods for the hadronic background rejection.
The remaining background can be estimated using also standard analysis methods, in particular using OFF region observed simultanously with the source of interest.

\begin{figure}[t]
\includegraphics[width=0.45\textwidth]{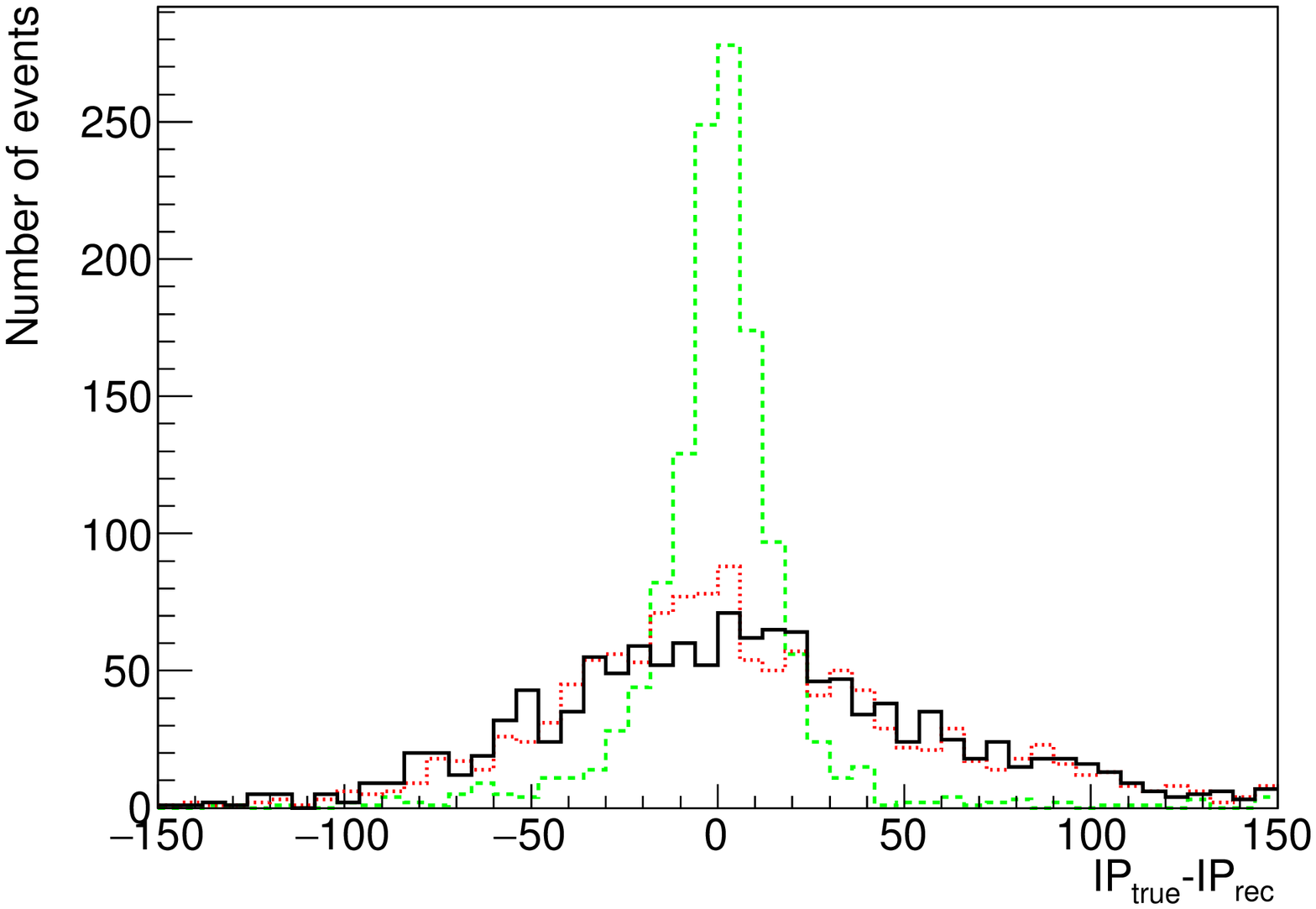}
\includegraphics[width=0.45\textwidth]{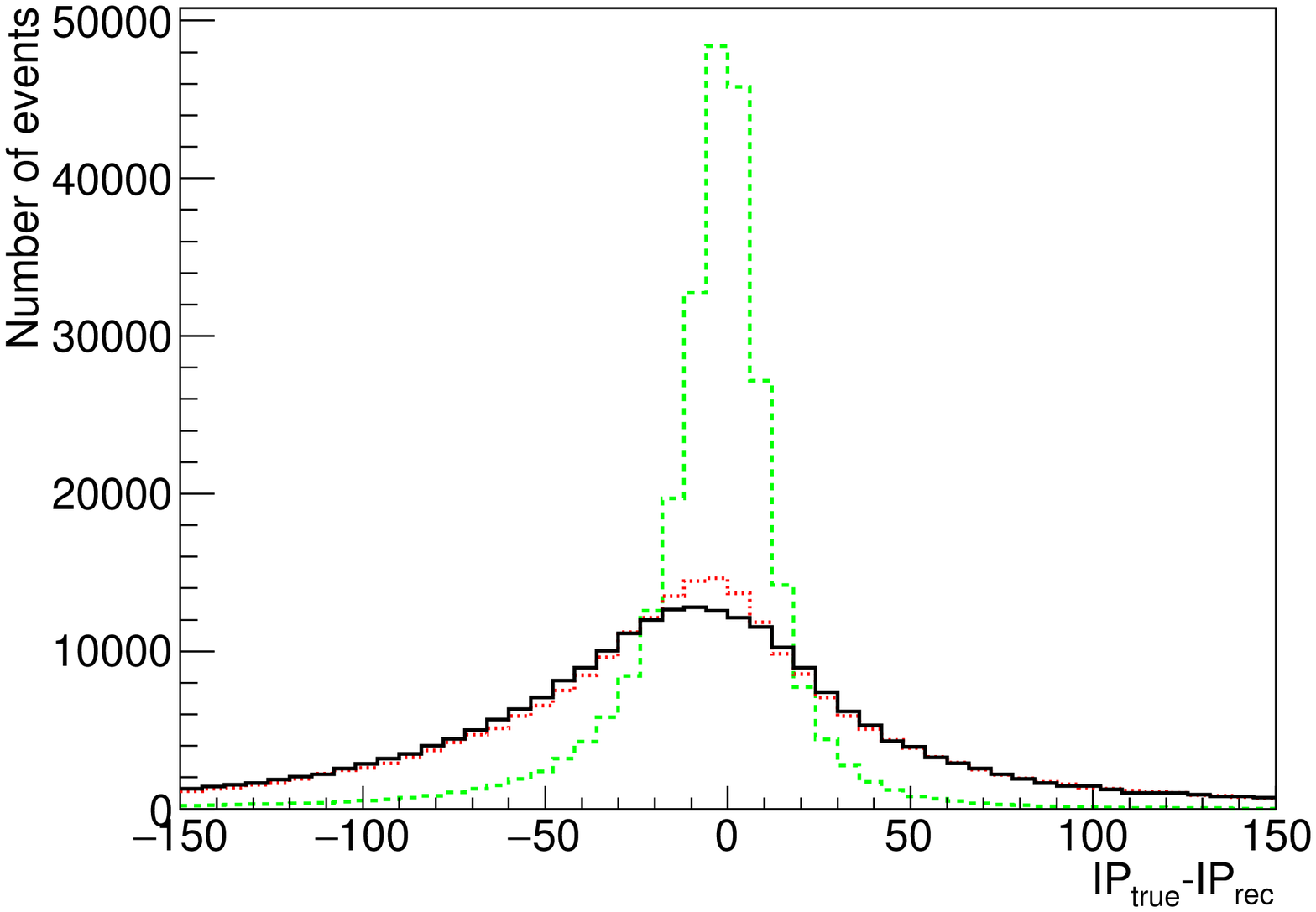}
\caption{
  Difference between the reconstructed and true impact parameter: for standard stereo reconstruction (black solid) and obtained from the fit of the toy model (red dotted fitting source position, green dashed assuming nominal source position).
  Only events with estimated energy between 30 and 300\,GeV are used.
  Top panel shows events with at least 2 V-shaped images, bottom panel shows the remaining events. 
}
\label{fig:imp_reco}
\end{figure}

We check also if the reconstruction of the V-shaped events can be improved using the toy model described in Section~\ref{toy:model}. 
As an example we perform a joint fit to all the images of an V-shaped event by fixing the $E_0$ parameter to the estimated energy of the shower. 
In the minimization procedure $x_0$ and $y_0$ are given the starting values obtained from the standard stereoscopic reconstruction and are left free to vary. The $H_1$ and $f_{e^{-}}$ are treated as nuisance parameters.
In Fig.~\ref{fig:imp_reco} we compare the impact reconstruction of the classical stereoscopic reconstruction with the one obtained from the toy model fitting procedure for V-shaped events.  
In the fitting procedure we either fix the arrival direction of the gamma ray to its true value, or let it free in the fitting procedure. 

Since the toy model fitting procedure is using prior knowledge of the arrival direction of the shower its resolution of the impact reconstruction is nominally much better (the RMS of the distribution is 34\,m compared to 71\,m). 
If the true direction of the shower is not known and let free to vary, in the toy model fitting procedure the obtained RMS of the distribution is 66\,m, much closer to the value obtained from the standard reconstruction. 
Applying to events without V-shaped images a similar procedure, fitting however instead of two separate subshowers a single shower moving along the nominal direction of the source allows us to obtain a similar improvement in the resolution of the impact parameter when exploiting the knowledge of the true source direction.
The obtained RMS of the distribution in this case is 77\,m for standard reconstruction, 76\,m for geometrical fitting when the source position is left free to vary and 31\,m when the source position is set to the nominal value. 

\subsection{Geometrical energy estimation of V-shaped events}
For each image that survived the selection criteria described in Section~\ref{sec:sel} we fit the toy model described in Section~\ref{toy:model} simultaneously to all images. 
We minimize using MINUIT2\footnote{\url{https://root.cern.ch/root/html534/guides/minuit2/Minuit2.pdf}} the sum of squared distances of all the pixels in all the telescopes surviving the cleaning to the closest (for a given pixel) of the two lines given by the toy model. 

\begin{figure}[t]
\includegraphics[width=0.49\textwidth]{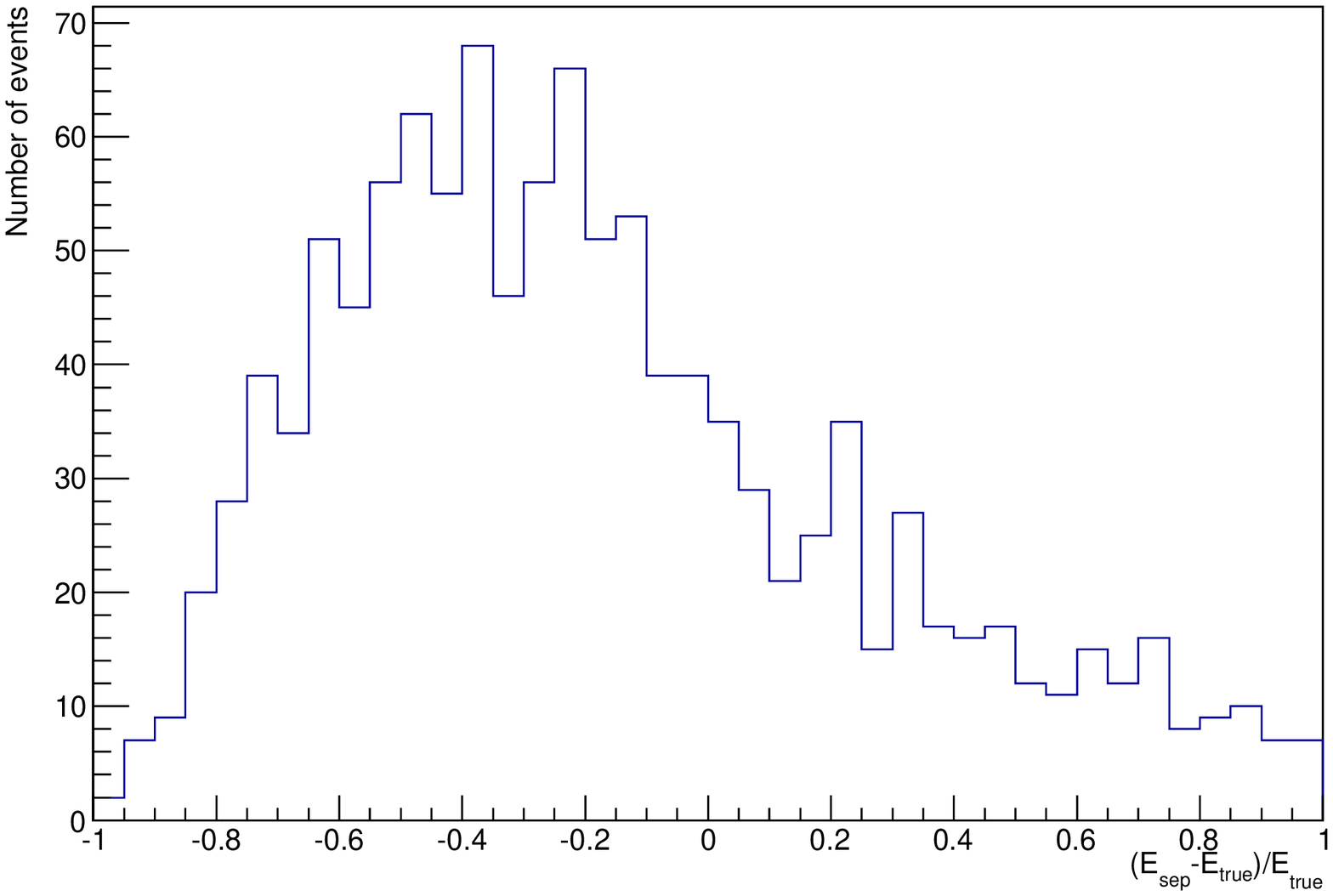}
\caption{Resolution of the energy estimation method using geometrical fit to V-shaped images.
  Only events with estimated (according to classical stereo reconstruction method) energy between 30 and 300\,GeV are used.
In the fitting procedure the true values of the source position, core location and height of the first interaction were used. 
}
\label{fig:corr}
\end{figure}

To test the method we fix the core location and height of the first interaction to the true values obtained from the MCs leaving free only the energy of the primary gamma ray and the energy division parameter $f_{e^{-}}$. 
In Fig.~\ref{fig:corr} we plot the correlation of the energy obtained from such a fitting procedure to the true energy of the event.
While there is a large spread, the correlation is visible between the true and fitted energy.
The energy resolution of the method calculated as the standard deviation of the distribution (computed in the range shown in Fig.~\ref{fig:corr}) is 43\%. 

\begin{figure}[t]
\includegraphics[width=0.49\textwidth]{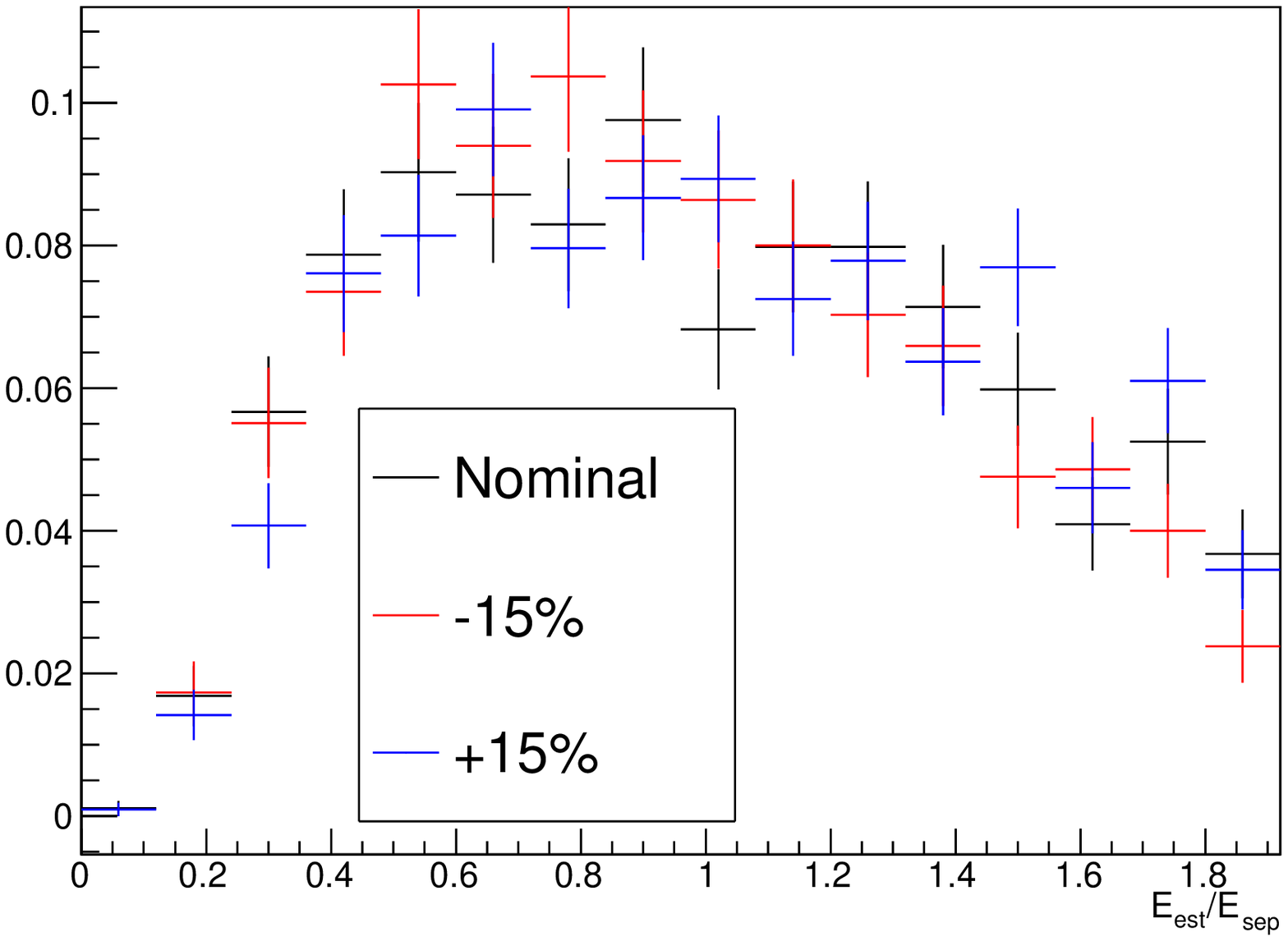}
\caption{Distribution of the ratio of the energy obtained from the classical stereoscopic reconstruction and image size ($E_{est}$) to the energy obtained in the image shape fit ($E_{sep}$).
  Black lines shows the results for the nominal light yield of the telescopes, while the total light throughput decreased or increased by 15\% is shown in red and blue respectively.
The energy range, event selection and image fitting procedure as in Fig.~\ref{fig:corr}.
}
\label{fig:eratio}
\end{figure}

To test if this alternative method of the energy estimation (purely geometrical, without using the total light yield seen by the telescope) can be exploited to validate the energy scale of Cherenkov telescope we make a distribution of the ratio of the energy estimation from the stereoscopic reconstruction ($E_{est}$) to the one obtained from the above-described fitting procedure ($E_{sep}$). 
We perform the study on three sets of MCs. 
The first, reference Set 1 has nominal total light throughput reflectivity values.
In the second and third we have artificially modified the light throughput by $\pm15\%$ to simulate a systematic uncertainty in the light scale of the telescopes.
Such a modification can simulate imperfections in the knowledge of the light collection parameters of the telescope (e.g. quantum efficiency of light sensors, reflectivity of the mirrors) or in the transmission of the atmosphere. 
All three sets are reconstructed using the nominal (Set 1) MCs. 
This causes a systematic bias in the energy estimation of Set 2 and Set 3. 
In Fig.~\ref{fig:eratio} we compare the distribution of $E_{est}/E_{sep}$ for the three sets of MCs. 
While the shift of the energy scale is partially reflected in the distribution, the effect is very small. 
The shift of the energy scale by $\pm15\%$ results in a shift by only $\pm 3\%$.
Such a shift is unfortunately too small for practical application of the method, which will be burdened by a number of its own systematic errors and statistical uncertainties (limited by the rate of V-shaped events and residual hadronic background). 
The fact that the shift of the $E_{est}/E_{sep}$ proxy is smaller than the shift of $E_{est}$ alone is most probably caused by the bias caused by the height of the first interaction discussed in Section~\ref{sec:epem}.
Consider a case in which the light throughput of the telescope is overestimated resulting in underestimation of $E_{est}$.
In such a case mostly the events that start high in the atmosphere and those that have larger GF separation will be lost as they do not produce enough light to survive the trigger and reconstruction.
Since the toy model cannot take into accout the trigger bias the average separations at a given energy given by the toy model are underestimated.
Therefore, the fitting procedure counteracts it by favouring lower energies.
As a result the effect on $E_{est}/E_{sep}$ is smaller than on $E_{est}$ alone. 

\section{Discussion and conclusions}\label{sec:con}
We investigated the propagation of the $e^+e^-$ pair produced in the first interaction of a $\gamma$ ray with atmospheric nucleus. 
In the energy range most interesting for the LST telescopes (30--300\,GeV) the deflection of the first $e^+e^-$ pair produces a shift of the two components by $\sim0.1^\circ$. 
If the first interaction occurred high in the atmosphere, and the impact factor of the shower is of the order of a few tens of meter the resulting image can have a very clear V-shaped signature. 
We derived a simple method to identify the V-shaped images candidates. 
We investigated how the \progname{Chimp/MARS} analysis chain of CTA reconstructs events which pass selection criteria for V-shaped images.
Despite non-regular shape, the angular resolution as well as energy resolution are similar to the ones obtained from regular, elliptical images.  
The gamma/hadron separation parameter is slightly shifted to more hadron-like values in case at least two images show V-shaped characteristics. 
The toy model fitting procedure can be used for more precise estimation of the impact parameter of V-shaped events, however the improvement is mostly related to the assumed nominal position of the gamma-ray source.
Similar improvement can be also achieved by applying a simplified geometrical fitting method to classical elliptical images for point-like sources by exploiting the knownledge of the nominal source position. 
The toy model fitting procedure can be used to perform an alternative estimate of the energy of V-shaped events, not dependent on the total light yield of the telescopes.
We investigated if the comparison of such an estimate of the energy with the classical one obtained from stereoscopic reconstruction of the shower can be used to validate the energy scale of the Cherenkov telescope.
The change of the telescope reflectivity leaves an imprint on the distribution of the ratio of both energy estimations.
However, in the presented analysis the effect has too small magnitude (most probably due to strong dependence of the deflection on the height of the first interaction) that limits the practical usage of the method in the presented here form.

The performance of the proposed here method might be different at larger zenith angles. The first interaction would happen at higher altitude, resulting in longer path of deflection of $e^+$ and $e^-$ before the development of their subcascades.
This would in turn produce more separated sub-images. 
The observations at higher zenith angle however result in stronger absorption of light in the atmosphere and effectively higher threshold.
This would enhance the observation bias on the distribution of the height of the first interaction seen in Fig.~\ref{fig:dcog} worsening the performance of the method.
Similarly, the MST telescopes can be used together with LSTs to reconstruct the two subshowers in a more precise way.
However, it should be noted that since in V-shaped images the $e^+$ and $e^-$ subcascades are separated on the camera, the energy threshold for observation of such events should be about a factor two larger then nominal energy threshold for classical elliptical images.
As the most pronounced V-shaped events occured for LSTs for energies about 100\,GeV, they would be at or below the analysis threshold for MSTs.
Finaly, the performance of the method might be improved by using a more refined image fitting algorithm, possibly using templates of the two subshowers (see e.g. \citealp{nr09, par14}). 

\section*{Acknowledgements}
This work is supported by the grant through the Polish Narodowe Centrum Nauki No. 2015/19/D/ST9/00616.
DS is supported by the Narodowe Centrum Nauki grant No. UMO-2016/22/M/ST9/00583. 
We would like to thank CTA Consortium and MAGIC Collaboration for allowing us to use their software. 

This paper has gone through internal review by the CTA Consortium.
In particular we would like to thank Abelardo Moralejo, Konrad Bernl\"ohr, Bruno Kh\'elifi, Jose Luis Contreras and Johan Bregeon for helpful discussions.

\section*{References}

\def\apjl{ApJL}
\def\apj{ApJ}
\def\aap{A\&A}
\def\procspie{Proceedings of the SPIE}


\end{document}